\documentclass[a4paper]{article}

\usepackage[english]{babel}
\usepackage[utf8x]{inputenc}
\usepackage[T1]{fontenc}
\usepackage{authblk}
\usepackage{blkarray}
\usepackage[normalem]{ulem}

\usepackage[a4paper,top=3cm,bottom=2cm,left=3cm,right=3cm,marginparwidth=1.75cm]{geometry}

\usepackage{amsmath}
\usepackage{graphicx}
\usepackage[colorlinks=true, allcolors=blue]{hyperref}
\usepackage{color}

\usepackage{bm}

\usepackage{fancyvrb} 

\definecolor{tomato}{RGB}{255,99,71}
\definecolor{lightblue}{RGB}{173,216,230}

\usepackage[natbibapa]{apacite}
\usepackage{float}
\usepackage{amsfonts}

\newcommand\scalemath[2]{\scalebox{#1}{\mbox{\ensuremath{\displaystyle #2}}}}

\usepackage{adjustbox}
\usepackage{array}
\newcolumntype{R}[2]{%
    >{\adjustbox{angle=#1,lap=\width-(#2)}\bgroup}%
    l%
    <{\egroup}%
}

\usepackage{mathabx}

\makeatletter
\def\blfootnote{\xdef\@thefnmark{}\@footnotetext}
\makeatother

\title{A Tutorial on Estimating Time-Varying Vector Autoregressive Models}

\author[1]{Jonas Haslbeck\thanks{Contact: \url{jonashaslbeck@gmail.com} | \url{www.jonashaslbeck.com}}}
\author[2]{Laura Bringmann}
\author[1]{Lourens Waldorp}

\date{}

\affil[1]{Psychological Methods Group, University of Amsterdam}
\affil[2]{Department of Psychometrics and Statistic, University of Groningen}

\begin{document}

\maketitle

\begin{abstract}
	Time series of individual subjects have become a common data type in psychological research. These data allow one to estimate models of within-subject dynamics, and thereby avoid the notorious problem of making within-subjects inferences from between-subjects data, and naturally address heterogeneity between subjects. A popular model for these data is the Vector Autoregressive (VAR) model, in which each variable is predicted as a linear function of all variables at previous time points. A key assumption of this model is that its parameters are constant (or stationary) across time. However, in many areas of psychological research time-varying parameters are plausible or even the subject of study. In this tutorial paper, we introduce methods to estimate time-varying VAR models based on splines and kernel-smoothing with/without regularization. We use simulations to evaluate the relative performance of all methods in scenarios typical in applied research, and discuss their strengths and weaknesses. Finally, we provide a step-by-step tutorial showing how to apply the discussed methods to an openly available time series of mood-related measurements.
\end{abstract}

\blfootnote{This article has been accepted for publication in Multivariate Behavioral Research, published by Taylor \& Francis.}

\section{Introduction}

The ubiquity of mobile devices has led to a surge in time series (or intensive longitudinal) data sets from single individuals \citep[e.g.,][]{bringmann2013network, kramer2014therapeutic, hartmann2015experience, kroeze2016personalized, van2017temporal, bak2016n, snippe2017impact, fisher2017exploring, groen2019capturing}. This is an exciting development because these data allow one to model within-subject dynamics, which avoids the notorious problem of making within-subjects inferences from between-subjects data, and naturally addresses heterogeneity between subjects \citep{fisher2018lack, molenaar2004manifesto}. The ability to analyze within-subjects data therefore promises to be a major leap forward both for psychological research and applications in (clinical) practice.

\begin{figure}[h]
	\label{fig_intro}
	\centering
	\includegraphics[width=.85\linewidth]{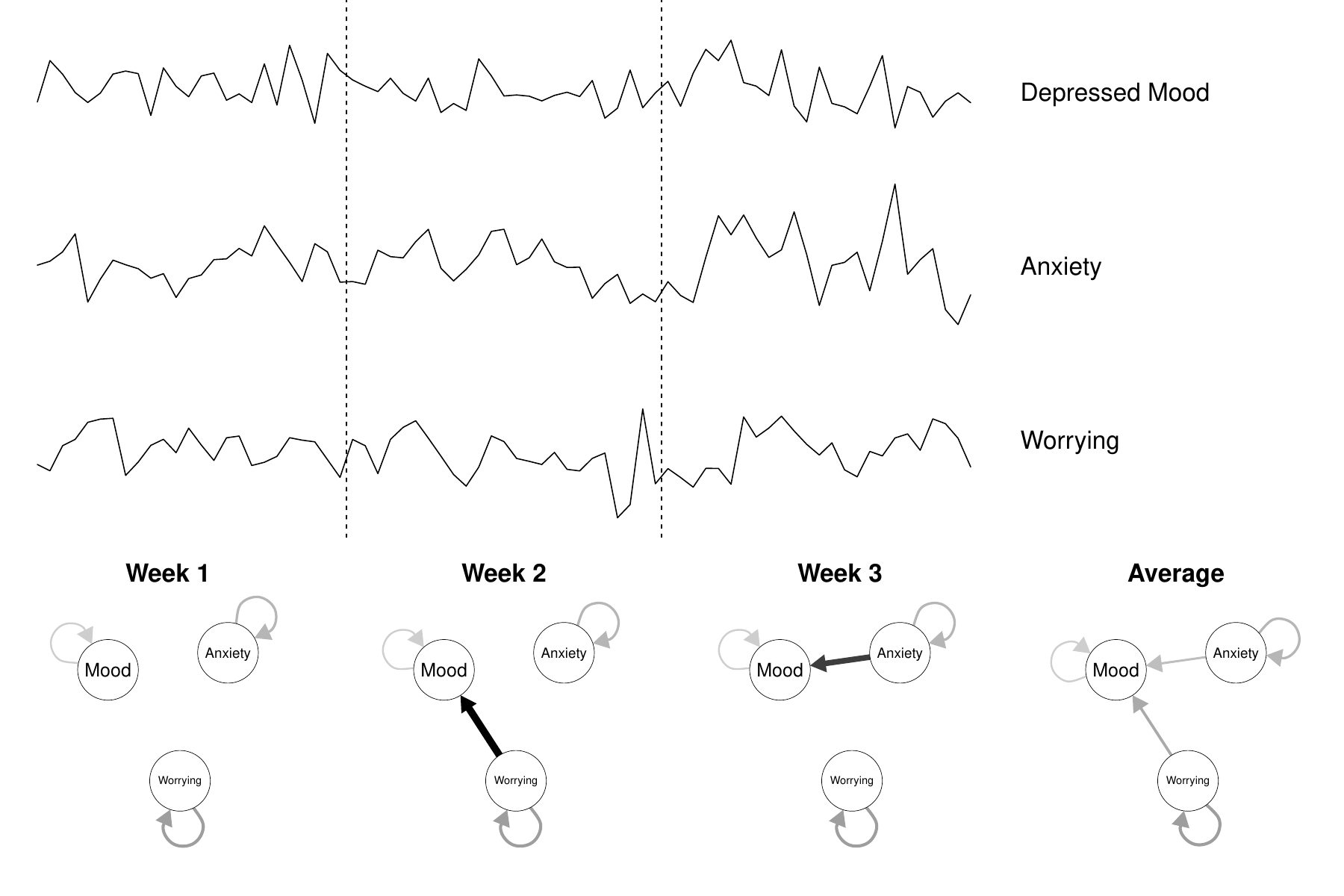}
	
	\caption{Upper panel: hypothetical repeated measurements of \textit{Depressed Mood}, \textit{Anxiety} and \textit{Worrying}, generated from a time-varying lag 1 VAR model. Lower panel: the time-varying VAR-model generating the data shown in the upper panel. It consists of three models, one for each week. The fourth model (left to right) indicates the average of the three models, which is what one obtains when estimating a stationary VAR model on the entire time series.}
\end{figure}

A key assumption of all standard time series models is that all parameters of the data generating model are constant (or stationary) across the measured time period. This is called the \textit{assumption of stationarity}\footnote{We use this definition of stationarity, because for VAR models with eigenvalues within the unit circle, which we focus on in this paper, it is equivalent to definitions based on the moments of distributions. This implies that we do not consider diverging VAR models (with eigenvalues outside the unit circle) which have a non-stationary distribution while its parameters are constant across time.}. While one often assumes constant parameters, changes of parameters over time are often plausible in psychological phenomena. As an example, take the repeated measurements of the variables \textit{Depressed Mood}, \textit{Anxiety} and \textit{Worrying}, modeled by a time-varying first-order Vector Autoregressive (VAR) model shown in Figure \ref{fig_intro}. In week 1, there are no cross-lagged effects between any of the three variables. However, in week 2 we observe a cross-lagged effect from \textit{Worrying} on \textit{Mood}. A possible explanation could be a physical illness in week 2 that moderates the two cross-lagged effects. In week 3, we observe a cross-lagged effect from \textit{Anxiety} on \textit{Mood}. Again, this could be due to an unobserved moderator like a stressful period at work. The fourth visualization shows the average of the previous three models, which is the model one would obtain by estimating a stationary VAR model on the entire time series. In this situation, the stationary model is clearly inappropriate because it is different to the true model across \textit{all} intervals of the time series. 

Time-varying models are of central interest when studying psychological phenomena from a within-person perspective. For example, in the network approach to psychopathology, it is suggested that mental disorders arise from causal interactions among symptoms \citep[see also][]{borsboom2013network, schmittmann2013deconstructing, reviewDon2019}. This means that the interactions between symptoms are different for healthy and unhealthy individuals \citep{pe2015emotion,van2015association} and that the interactions in an individual change when she or he transitions from a healthy to an unhealthy state (or vice versa). Time-varying models are able to capture this change. Next to detecting these changes, they may also shed light on why those changes occurred. For example, one could correlate time-varying parameters with contextual factors such as elevated stress levels, social setting or major life events and thereby possibly uncover conditions and events that predict the onset of mental disorders. Time-varying models can also be used to study how parameters change in response to interventions. For example, in Section \ref{sec_tutorial} we will fit a time-varying VAR model on ESM measurements during a double-blind medication reduction study \citep{wichers2016critical}.

Time-varying models are also central to the idea of Early Warning Signals \citep[EWS;][]{scheffer2009early}. For example, \cite{wichers2016critical} suggested to anticipate phase-transitions between healthy and unhealthy states with EWS such as time-varying autocorrelation and variance \citep[see also][]{van2014critical}. Time-varying VAR models are an extension of these EWS to multivariate time-series. Anticipating the sensitive periods around phase transitions is interesting, because during those periods treatment may be more efficient \citep{olthof2019critical}. This means that time-varying models could be used as a tool to monitor patients and determine periods during which treatment is most promising.

In this tutorial paper we provide an introduction to how to estimate a time-varying version of the Vector Autoregressive (VAR) model, which is arguably the simplest multivariate time series model for temporal dependencies in continuous data, and is used in many of the papers cited above. We will focus on two sets of methods recently proposed by the authors to estimate such time-varying VAR models: \cite{bringmann2018modeling} presented a method based on splines using the Generalized Additive Modeling (GAM) framework, which estimates time-varying parameters by modeling them as a spline function of time; and \cite{haslbeck2015mgm} suggested a method based on penalized kernel-smoothing (KS), which estimates time-varying parameters by combining the estimates of several local models spanning the entire time series. While both methods are available to applied researchers, it is unclear how well they and their variants (with/without regularization or significance testing) perform in situations that are typical in applied research. We aim to improve this situation by making the following contributions:

	\begin{enumerate}
		\item We report the performance of GAM based methods with and without significance testing, and the performance of KS based methods with and without regularization in situations that are typical for Experience Sampling Method (ESM) studies.
        \item We discuss the strengths and weaknesses of all methods and provide practical recommendations for applied researchers
		\item We compare time-varying methods to their corresponding. stationary counterparts to address the question of how many observations are necessary to identify the time-varying nature of parameters.
		\item We provide tutorials on how to estimate time-varying VAR models using both methods on an openly available intensive longitudinal dataset using the R-packages \emph{mgm} and \emph{tvvarGAM}.
	\end{enumerate}

The paper is structured as follows. In Section \ref{methods_VAR} we define time-varying VAR models, which are the focus of this paper. We next present two sets of methods to recover such models: one method based on splines with and without significance testing (Section \ref{methods_GAM}), and one method based on kernel estimation with and without regularization (Section \ref{methods_ks}). In Sections \ref{sim_A_randomgraph} and \ref{sim_B_UT} we report two simulation studies that investigate the performance of these two models and their stationary counterparts. In Section \ref{sec_tutorial} we provide a fully reproducible tutorial on how to estimate a time-varying VAR model from an openly available time series data set collected in ESM studies using the kernel smoothing method using the R-package \emph{mgm} (we repeat the same tutorial with the GAM method in the appendix). Finally, in Section \ref{sec_discussion} we discuss possible future directions for research on time-varying VAR models.

\section{Estimating Time-Varying VAR Models}\label{sec_estimationmethods}

We first introduce the notation for the stationary first-order VAR model and its time-varying extension (Section \ref{methods_VAR}) and then present the two methods for estimating time-varying VAR models: the GAM-based method (Section \ref{methods_GAM}) and the penalized kernel-smoothing-based method (Section \ref{methods_ks}). We discuss implementations of related methods in Section \ref{methods_related}.

\subsection{Vector Autoregressive (VAR) Model}\label{methods_VAR}

In the first-order Vector Autoregressive (VAR(1)) model, each variable at time point $t$ is predicted by all variables (including itself) at time point $t-1$. Next to a set of intercept parameters, the VAR(1) model is comprised by autoregressive effects, which indicate how much a variable is predicted by itself at the previous time point, and cross-lagged effects, which indicate how much a variable is predicted by all other variables at the previous time point.

Formally, the variables $\mathbf{X}_t \in \mathbb{R}^p$ at time point $t \in \mathbb{Z}$ are modeled as a linear combination of the same variables at $t-1$

\begin{equation}\label{eq_VAR}
\mathbf{X}_t 	=  \bm{\beta_0} + \bm{B} \mathbf{X}_{t-1} + \bm{\varepsilon} =  
\begin{bmatrix}
X_{t, 1}  \\ 
\vdots \\
X_{t, p}           
\end{bmatrix}
= 
\begin{bmatrix}
\beta_{0, 1}  \\ 
\vdots \\
\beta_{0, p}           
\end{bmatrix}
+
\begin{bmatrix}
\beta_{1, 1} & \dots & \beta_{1, p}           \\
\vdots & \ddots           & \vdots  \\
\beta_{p, 1}           & \dots & \beta_{p, p}
\end{bmatrix}
\begin{bmatrix}
X_{t-1, 1}  \\ 
\vdots \\
X_{t-1, p}           
\end{bmatrix}
+ 
\begin{bmatrix}
\epsilon_1  \\ 
\vdots \\
\epsilon_p           
\end{bmatrix},
\end{equation}

\noindent
where $\beta_{0,1}$ is the intercept of variable 1, $\beta_{1, 1}$ is the autoregressive effect of $X_{t-1, 1}$ on $X_{t, 1}$, and $\beta_{p, 1}$ is the cross-lagged effect of $X_{t-1, 1}$ on $X_{t, p}$, and we assume that $\bm{\varepsilon} = \{\epsilon_1, \dots, \epsilon_p \}$ are independent (across time points) samples drawn from a multivariate Gaussian distribution with variance-covariance matrix $\Sigma$. In this paper we do not model $\Sigma$, however, it can be obtained from the residuals of the model and used to estimate the inverse covariance matrix \citep[see e.g.,][]{epskamp2018gaussian}.

Throughout the paper we deal with first-order VAR models in which all variables at time point $t$ are a linear function of all variables at time point $t-1$. In the interest of brevity we will therefore refer to this first-order VAR model (or VAR(1) model) as a VAR model. More lags can be included by adding further parameter matrices and lagged variable vectors $\bm{X}_{t-k}$ (for a lag of $k$) to the model in (\ref{eq_VAR}). Note that while we focus on VAR(1) models in the this paper, the presented methods can be used to estimate time-varying VAR models with any set of lags. For a detailed description of VAR models we refer the reader to \cite{hamilton1994time}.

In both the GAM and the KS method we estimate (\ref{eq_VAR}) by predicting each of the variables $X_{t, i}$ for $i \in \{1, \dots, p\}$ separately. Specifically, we model 

\begin{equation}\label{eq_conditional}
X_{t, i} = 
\beta_{0, i}
+ 
\bm{\beta_i}
\mathbf{X}_{t-1}
+
\epsilon_i
=
\beta_{0, i}
+ 
\begin{bmatrix}
\beta_{i,1} & \dots & \beta_{i,p}
\end{bmatrix}
\begin{bmatrix}
X_{t-1, 1}  \\ 
\vdots \\
X_{t-1, p}           
\end{bmatrix}
+
\epsilon_i
,
\end{equation}

\noindent
for all $i \in \{1, \dots, p\}$, where $\bm{\beta_i}$ is the $1 \times p$ vector containing the lagged effects on $X_{t, i}$. After estimating the parameters in each equation, we combine all estimates to the VAR model in (\ref{eq_VAR}).

In order to turn the stationary VAR model in (\ref{eq_VAR}) into a time-varying VAR model, we introduce a time index for the parameter matrices

\begin{equation}\label{eq_VAR_timev}
\mathbf{X}_t 	=  \bm{\beta_{0, t}} + \bm{B_t} \mathbf{X}_{t-1} + \bm{\varepsilon}
\;
.
\end{equation}

This allows a different parameterization of the VAR model at \emph{each time point} and thereby allows the model to vary across time. Throughout this paper we assume that the time-varying parameters are smooth deterministic functions of time. We define a smooth function as a function for which the first derivative exists everywhere. In the following two subsections we introduce two different ways to estimate such a time-varying VAR model.

The VAR model has often been discussed and visualized as a network model \citep{epskamp2018gaussian}, and also here we will use both statistical and network/graph terminology. To avoid confusion between the two terminologies, we explicitly state how the terms in the two terminologies correspond to each other. From the statistical perspective there are two types of lagged effects between pairs of variables: autocorrelations (e.g., $X_{t-1} \rightarrow X_t$) and cross-lagged effects (e.g., $X_{t-1} \rightarrow Y_t$). In the network terminology variables are nodes, and lagged effects are represented by directed edges. An edge from a given node on itself is also called a self-loop, and represents autocorrelation effects. The value of lagged effects is represented in sign and the absolute value of the edge-weights of the directed edges. If an edge-weight between variables $X_t$ and $Y_{t-1}$ is nonzero, we say that the edge from $X_t$ and $Y_{t-1}$ is present. \emph{Sparsity} refers to how strongly connected a network is: if many edges are present, sparsity is low; if only few edges are present, sparsity is high. On a node-level, sparsity is captured by the \emph{indegree} (how many edges point towards a node) and \emph{outdegree} (how many edges point away from a node). In statistical terminology indegree is the number of incoming lagged effects on variable $X$, and outdegree the number outgoing lagged effects from variable $X$.

\subsection{The GAM Method}\label{methods_GAM}

In this section we explain how to estimate a time-varying VAR model using the Generalized Additive Model (GAM) framework, which allows for non-linear relationships between variables \cite[see also][]{bringmann2017changing,bringmann2018modeling}. We leverage the GAM framework for the estimation for time-varying models by using it to define each parameter as a function of time. Because GAMs are able to represent non-linear functions, this allows us to recover non-linear time-varying parameters. In what follows we illustrate how this approach works for the simplest possible example, a model consisting only of a time-varying intercept parameter, $y=\beta_{0,t} + \varepsilon$.

\begin{figure}[h]
\centering
\includegraphics[width=1\linewidth]{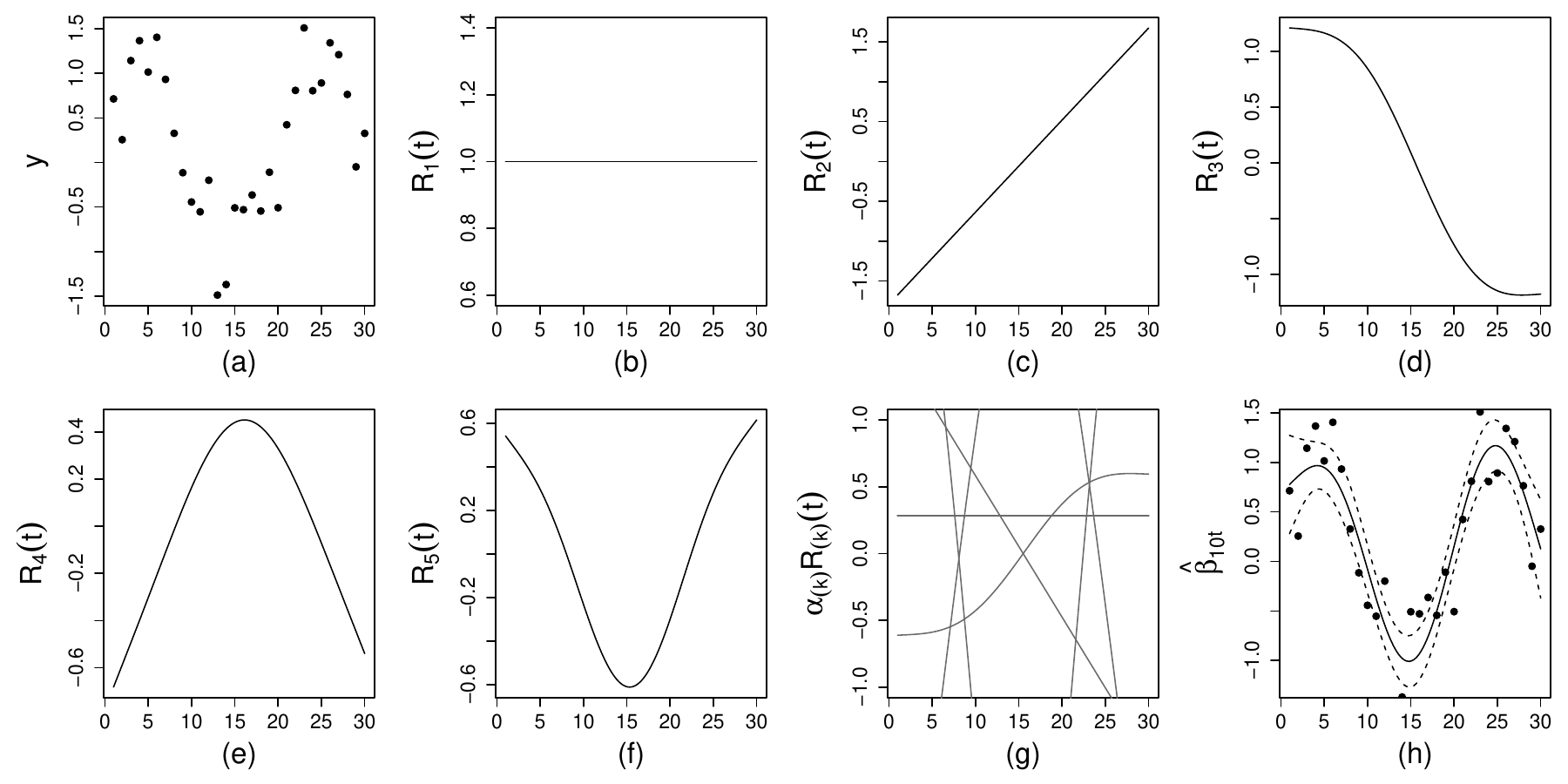}
	\caption{An example of the basis function for a time-varying parameter $\hat{\beta}_{0,t}$. In panel (a) the data are shown. In panel (b)-(f) the estimated 5 basis functions are given and panel (g) shows the weighted basis functions. In the last panel (h) the final smooth function is illustrated with credible intervals around the smooth function.
	}	\label{fig_thinplate}
\end{figure}

Panel (a) of Figure \ref{fig_thinplate} shows that the values of $y$ are varying over time, so the intercept will have to be time-varying as well, if the intercept-only model is supposed to fit the data well. This is achieved by summing the following five basis functions

\begin{equation}\label{eq:arc1}
\hat{\beta}_{0,t} = \hat{\alpha}_1 R_1(t) + \hat{\alpha}_2 R_2(t) + \hat{\alpha}_3 R_3(t) +   \hat{\alpha}_4 R_4(t) +  \hat{\alpha}_5 R_5(t),
\end{equation} 

\noindent
which are displayed in panels (b) - (f) in Figure \ref{fig_thinplate}. Panel (g) overlays all used basis functions, and panel (h) displays the estimate of the final smooth function $\hat{\beta}_{0,t}$, which is obtained by adding up the weighted basis functions ($\hat{\alpha}$) (see panel (g) and (h) of Figure \ref{fig_thinplate}). The optimal regression weights are estimated using standard linear regression techniques. The same rationale is applied to every time-varying parameter in the model. 

There are several different spline bases such as cubic, P-splines, B-splines, and thin plate splines. The advantage of thin plate splines, which is the basis used here, is that one does not have to specify knot locations, resulting therefore in fewer subjective decisions that need to be made by the researcher \citep{wood2006generalized}. The basis functions in Figure \ref{fig_thinplate} exemplify the thin plate spline basis. In the figure, panels (b)-(f) show that each additional basis function ($R$) increases the nonlinearity of the final smooth function. This is reflected in the fact that every extra basis function is more ``wiggly'' than the previous basis functions. For example, the last basis function in panel (f) is ``wigglier'' than the first basis function in panel (b). The spline functions used here are smooth up to the second derivative. Thus, a key assumption of the GAM method is that all true time-varying parameter functions are smooth as well. This assumption is also called the assumption of \textit{local stationarity}, because smoothness implies that the parameter values that are close in time are very similar, and therefore locally stationary. This would be violated by, for example, a step function, where the GAM method would provide incorrect estimates around a ``jump'' (but would still give good estimates for the two constant parts). 

As the number of basis functions determines the nonlinearity of the smooth function (e.g., $\hat{\beta}_{0,t}$), a key problem is how to choose the optimal number of basis functions. The final curve should be flexible enough to be able to recover the true model, but not too flexible as this may lead to overfitting \citep{andersen2009nonparametric,keele2008semiparametric}. The method used here to find the optimal number of basis functions is penalized likelihood estimation \citep{wood2006generalized}. Instead of trying to select the optimal number of basis functions directly, one can simply start by including more basis functions than would be normally expected, and then adjust for too much wiggliness with a \textit{wiggliness penalty}  \citep{wood2006generalized}. 

Thus, the problem of selecting the right number of basis functions is reduced to selecting the right wiggliness penalty. This is achieved using generalized cross-validation \citep{golub1979generalized}, where the penalty parameter with the lowest Generalized Cross-Validation (GCV) value is expected to give a good bias-variance trade-off. Specifically, the penalization decreases the influence of the basis functions ($R$) by reducing the values of their regression coefficients ($\hat{\alpha}$). Therefore, smoothness is imposed on the curve both through the choice of the number of basis functions and the final level of penalization on these basis functions.

To estimate time-varying VAR models with the GAM method, we use the \textit{tvvarGAM} package in \textit{R} \citep{tvvarGAM}, which is a wrapper around the \textit{mgcv} package \citep{wood2006generalized}. As the wiggliness penalty is automatically determined, the user only needs to specify a large enough number of basis functions. The default settings are the thin plate regression spline basis and 10 basis functions, which although an arbitrary number, is often sufficient \cite[see the simulation results in][]{bringmann2017changing}. The minimum number is in most models three basis functions. In general, it is recommended to increase the number of basis functions if it is close to the effective degrees of freedom (edf) selected by the model. The effective degrees of freedom is a measure of nonlinearity. A linear function has an edf of one, and higher edf values indicate wigglier smooth functions \citep{shadish2014using}. 

The GAM function in the \textit{mgcv} package outputs the final smooth function, the GCV value and the edf.  Furthermore,  the uncertainty of the smooth function is estimated with 95$\%$ Bayesian credible intervals \citep{wood2006generalized}. In the remainder of this manuscript we refer to this method as the GAM method. We refer to a variant of the GAM method, in which we set those parameters to zero whose 95$\%$ Bayesian credible interval overlaps with zero, with GAM(st), for ``significance thresholded''. With GLM we refer to the standard unregularized VAR estimator.

After the model is estimated, it is informative to check if the smooth functions were significantly different from zero (at some point over the whole time range), and if each smooth function had enough basis functions. Significance can be examined using the \textit{p}-values of each specific smooth function, which indicates whether the smooth function is significantly different from zero. To see whether there are enough basis functions, the edf of each smooth function can be examined. The edf value should be well below the maximum possible edf or the number of basis functions for the smooth function (or term) of interest \cite[in our case 10, see][]{wood2006generalized}. When the edf turns out to be too high, the model should be refitted with a larger (e.g., double) number of basis functions.

\subsection{The Kernel-smoothing Method}\label{methods_ks}

\begin{figure}[h]
	\begin{minipage}[b]{0.47\linewidth}
	\centering
	\includegraphics[width=1\linewidth]{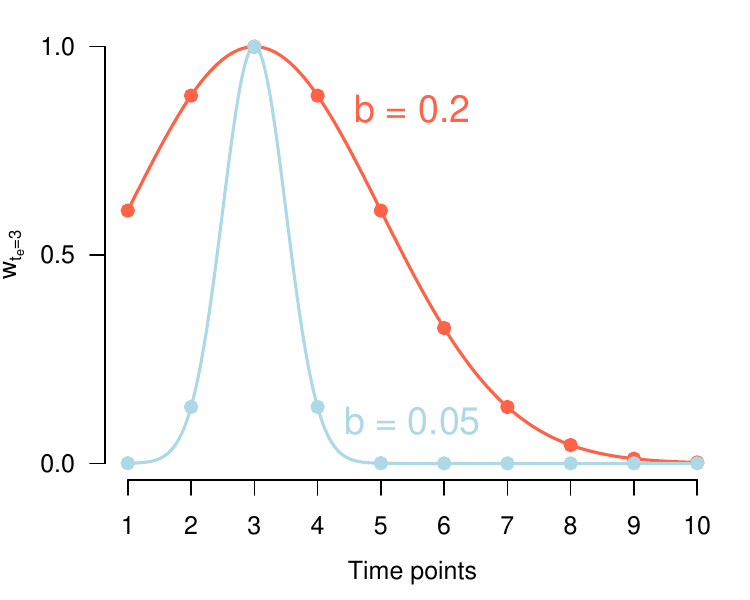}
	\end{minipage}
	\begin{minipage}[b]{0.53\linewidth}
			$$
			\scalemath{.87}{
				\bordermatrix{ \text{Time}   &X_{t, 1} & X_{t-1, 1} & \dots & X_{t, p} & \textcolor{tomato}{w_{ t_{e}=3}} & \textcolor{lightblue}{w_{ t^*_{e}=3}} \cr
					1     & 0.03    &  -0.97  & \dots    & -0.08 & \textcolor{tomato}{0.61} & \textcolor{lightblue}{0.00}    \cr
					2     & 1.15    &  -1.07  & \dots    & -0.56 & \textcolor{tomato}{0.88} & \textcolor{lightblue}{0.14}    \cr
					3     & 0.11    &  0.63   & \dots    & 1.09  & \textcolor{tomato}{1.00} & \textcolor{lightblue}{1.00} \cr
					4     & -1.08   &  0.13   & \dots    & 1.88  & \textcolor{tomato}{0.88} & \textcolor{lightblue}{0.14} \cr
					5     & -0.93   &  1.00   & \dots    & -0.29 & \textcolor{tomato}{0.61} & \textcolor{lightblue}{0.00} \cr
					6     & -1.08   &  0.17   & \dots    & -1.36 & \textcolor{tomato}{0.32} & \textcolor{lightblue}{0.00} \cr
					7     & 0.27    &  -1.72  & \dots    & -1.13 & \textcolor{tomato}{0.14} & \textcolor{lightblue}{0.00} \cr	
					8     & 0.03    &  -1.26  & \dots    & -0.97 & \textcolor{tomato}{0.04} & \textcolor{lightblue}{0.00} \cr
				    9     & -1.29   &  -1.05  & \dots    & -0.10 & \textcolor{tomato}{0.01} & \textcolor{lightblue}{0.00} \cr
					10     & -0.07     &  -0.04  & 1.05    & -0.12 & \textcolor{tomato}{0.00} & \textcolor{lightblue}{0.00} }
			}
			$$
			\vspace{.9cm}
			
	\end{minipage}
		
	\caption{Illustration of the weights defined to estimate the model at time point $t_e = 3$. Left panel: a kernel function defines a weight for each time point in the time series. Right panel: the weights shown together with the VAR design matrix constructed to predict $X_{t, 1}$.}	\label{fig_kernelsmooth}
\end{figure}

In the kernel-smoothing method one obtains time-varying parameters by estimating and combining a sequence of local models at different time points across the time series. A local model is estimated by weighting all observations depending on how close they are to the time point at which the local model is estimated. In Figure  \ref{fig_kernelsmooth} we show an example in which a single local model is estimated at time point $t_{e} = 3$. We do this by giving the time points close to $t_{e}$ a high weight and time points far away from $t_{e}$ a very small or zero weight. If we estimate models like this on a sequence of equally spaced estimation points across the whole time series and take all estimates together, we obtain a time-varying model.

Specifically, we use a Gaussian kernel $\mathcal{N}(\mu = t_{e}, \sigma^2 = b^2)$ function to define a weight for each time point in the time series

\begin{equation}\label{eq_kernelfunction}
	w_{j, t_{e}} = \frac{1}{\sqrt{2\pi b^2} } \exp 
	\left \{  
	- \frac{(j - t_e)^2}{2 b^2}
	\right  \}
	,
\end{equation}

\noindent
where $j \in \{1, 2, \dots, n \}$, which is the local constant or Nadaraya-Watson estimator \citep{fan1996applications}.

For the example shown in Figure \ref{fig_kernelsmooth} this means that the time point $t_{e} = 3$ gets the highest weight, and if the distance to $t_{e}$ increases, the weight becomes exponentially smaller. The same idea is represented in the data matrix in the right panel of Figure \ref{fig_kernelsmooth}: each time point in the multivariate time series is associated with a weight defined by the kernel function. The smaller we choose the bandwidth $b$ of the kernel function, the smaller the number of observations we combine in order to estimate the model at $t_e$: when using a kernel with bandwidth $b = 0.2$ (red curve), we combine more observations than when using the kernel with $b = 0.05$ (blue curve). The smaller the bandwidth the larger the sensitivity to detect changes in parameters over time. However, a small bandwidth means that less data is used and therefore the estimates are less reliable (e.g., only three time points when $b = 0.05$; see right panel of Figure  \ref{fig_kernelsmooth}).

Since we combine observations close in time to be able to estimate a local model, we have to assume that the models close in time are also similar. This is equivalent to assuming that the true time-varying parameter functions are smooth, or locally stationary. Thus, the key assumption of the kernel-smoothing approach is the same as in the spline approach. For the kernel-smoothing method, we need the additional assumption that the chosen bandwidth is small enough to capture the time-varying nature of the true model. For example, if the parameters of the true model vary widely over time, but the bandwidth is so large that at any estimation point almost the entire time series is used for estimation, it is impossible to recover the true time-varying function.

The weights $w_{j, t_{e}} $ defined in (\ref{eq_kernelfunction}) enter the loss function of the $\ell_1$-regularized regression problem we use to estimate each of the $p$ time-varying versions of the model in (\ref{eq_conditional})

\begin{equation}\label{KS_loss_function}
\hat{\bm{\beta_{t_{e}} }} = 
	\arg_{\bm{\beta_{t_{e}}, \beta_{0,t_{e}}}} \min 
	\left \{     
	 \frac{1}{n} 
	 \sum_{j = 2}^{n} w_{j, t_{e}}  
	 (X_{i, j} - \beta_{0,t_{e}} - \bm{\beta_{t_{e}} } \bm{X_{j-1}})^2 + \lambda_i ||  \bm{\beta}_t ||_1 
	\right \}
	,
\end{equation}

\noindent
where $X_{i, j}$ is the $j^\text{th}$ time point of the  $i^\text{th}$ variable in the design matrix, $||  \bm{\beta}_{t_{e}} ||_1 = \sum_{i=1}^p \sqrt{\beta_{i, t_e}^2} $ is the $\ell_1$-norm of $\bm{\beta_{t_{e}}}$, and $\lambda_i$ is a parameter controlling the strength of the penalty. Note that the indices $i$ and $t_e$ are fixed in (\ref{KS_loss_function}) because we estimate the time-varying VAR model equation by equation, separately for each estimation point $t_e$.

For each of the $p$ regressions, we select the $\lambda_i$ that minimizes the out-of-sample deviance in 10-fold cross validation \citep{friedman2010regularization}. In order to select an appropriate bandwidth $b$, we choose the $\hat{b}$ that minimizes the out of sample deviance \textit{across} the $p$ regressions in a time stratified cross validation scheme (for details see Section \ref{sim_A_estimation}). We choose a constant bandwidth for all regressions so we have a constant bandwidth for estimating the whole VAR model. Otherwise the sensitivity to detect time-varying parameters and the trade-off between false positives and false negatives differs between parameters, which is undesirable.

In $\ell_1$-penalized (LASSO) regression the squared loss is minimized together with the $\ell_1$-norm of the parameter vector. This leads to a trade-off between fitting the data (minimizing squared loss) and keeping the size of the fitted parameters small (minimizing $\ell_1$-norm). Minimizing both together leads to small estimates being set to exactly zero, which is convenient for interpretation. When using $\ell_1$-penalized regression, we assume that the true model is sparse, which means that only a small number of parameters $k$ in the true model are nonzero. If this assumption is violated, the largest true parameters will still be present, but small true parameters will be incorrectly set to zero. However, if we keep the number of parameters constant and let $n \rightarrow \infty$, $\ell_1$-regularized regression also recovers the true model if the true model is not sparse. For an excellent treatment on $\ell_1$-regularized regression see \cite{hastie2015statistical}.

As noted above, the larger the bandwidth $b$, the more data is used to estimate the model around a particular estimation point. Indeed, the data used for estimation is proportional to the area under the kernel function or the sum of the weights  $N_{\mathrm{util}} = \sum_{j=1}^{n} w_{j, t_{e}} $. Notice that $N_{\mathrm{util}}$ is smaller at the beginning and end of the time series than in the center, because the kernel function is truncated. This necessarily leads to a smaller sensitivity to detect effects at the beginning and the end of the time series. For a more detailed description of the kernel smoothing approach see \cite{haslbeck2015mgm}. In the remainder of this manuscript we refer to this method as KS(L1). With GLM(L1) we refer to the stationary $\ell_1$-penalized estimator.

\subsection{Related methods}\label{methods_related}

Several implementations of related models are available as free software packages. The R-package \textit{earlywarnings} \citep{earywarningsRpackage} implements the estimation of a time-varying AR model using a moving window approach. The R-package \textit{MARSS} \citep{MARSSpkg, MARSSpaper} implements the estimation of (time-varying) state-space models, of which the time-varying VAR model is a special case. While the state-space model framework is very powerful due to its generality, it requires the user to specify the way parameters are allowed to vary over time, for which often no prior theory exists in practice \citep{belsley1973time, tarvainen2004estimation}. In parallel efforts \cite{tvReg} developed the R-package \textit{tvReg}, which estimates time-varying AR and VAR models, as well as IRF, LM and SURE models, using kernel smoothing similar to the kernel smoothing approach described in the present paper, however does not offer $\ell_1$-regularization. Furthermore, the R-package \emph{bvarsv} \citep{bvarsv_package} allows one to estimate time-varying VAR models in a Bayesian framework. 

The R-package \emph{dynr} \citep{dynr} provides an implementation for estimating regime switching discrete time VAR models, and the R-package \emph{tsDyn} \citep{tsDyn} allows to estimate the regime switching Threshold VAR model \citep{tong1980threshold, hamaker2010regime}. These two methods estimate time-varying models that switch between piece-wise constant regimes, which is different to the methods presented in this paper, which assume that parameters change smoothly over time.

Another interesting way to modeling time-varying parameters is by using the fused lasso \citep{hastie2015statistical}. However, to our best knowledge this method is currently only implemented for the estimation of Gaussian Graphical Models: \cite{monti_2014} provide a Python implementation of the SINGLE algorithm \citep{monti_estimating_2014}, and \citep{GraphTime2017} provide a Python implementation of the (group) fused-lasso based method as presented in \cite{gibberd2017regularized}.

\section{Evaluating Performance via Simulation}\label{sec_simulation}

In this section we use two simulations to evaluate the performance of the methods introduced in Section \ref{sec_estimationmethods} in estimating time-varying VAR models. In the first simulation (Section \ref{sim_A_randomgraph}) we generate time-varying VAR models based on a random graph with fixed sparsity, which is the natural choice in the absence of any knowledge about the structure of VAR models in a given application. This simulation allows us to get a rough overview of the performance of all methods and their strengths and weaknesses. In the second simulation (Section \ref{sim_B_UT}), we generate time-varying VAR models in which we vary the level of sparsity. This simulation allows us to discuss the strengths and weaknesses of all methods in more detail, specifically, we can discuss in which situations methods with/without regularization or thresholding perform better. Finally, in Section \ref{sim_OverallDiscussion} we discuss the combined results of both simulations, and provide recommendations for applied researchers.

\subsection{Simulation A: Random Graph}\label{sim_A_randomgraph} 

In this simulation we evaluate the performance of all methods in estimating time-varying VAR models that are generated based on a random graph. We first describe how we generate these time-varying VAR models (Section \ref{sim_A_datagen}), discuss details about the estimation methods (Section \ref{sim_A_estimation}), report the results (Section \ref{sim_A_results}), and provide a preliminary discussion (Section \ref{sim_A_discussion}).

\subsubsection{Data generation}\label{sim_A_datagen}

We generated time-varying VAR models by first selecting the structure of a stationary VAR model and then turning this stationary VAR model into a time-varying one. Specifically, we used the following procedure to specify whether a parameter in the time-varying VAR(1) model is nonzero: we choose all our VAR models to have $p=10$ variables, which is roughly the number of variables measured in typical ESM studies. We start out with an empty $p \times p$ VAR parameter matrix. In this matrix we set all $p$ autocorrelations to be nonzero, since autocorrelations are expected to be present for most phenomena and are observed in essentially any application \citep[e.g., ][]{aan2012mood, snippe2017impact, wigman2015exploring}. Next, we randomly set $26$ of the $p \times p - p = 90$ off-diagonal elements (the cross-lagged effects) to be present. This corresponds to an edge probability of $P(\text{edge}) \approx 0.29\;$\footnote{We set a fixed number of elements to nonzero instead of using draws with $P(\mathrm{edge}) = 0.2$, because we resample the VAR matrix until it represents a stable VAR model (the absolute value of all eigenvalues is smaller than $1$). By fixing the number of nonzero elements we avoid biasing $P(\mathrm{edge})$ through this resampling process. Thus, all VAR matrices in each iteration and at each time point has no eigenvalue with absolute value greater than 1.}. This approach returns an initial $p \times p$ matrix with ones in the diagonal and zeros and ones in the off-diagonal. 

In a second step we use the structure of this VAR model to generate a time-varying VAR model. Specifically, we randomly assign to each of the nonzero parameters one of the sequences (a) - (g) in Figure \ref{fig_data_gen_8types}:

\begin{figure}[H]
    \centering
    \includegraphics[width=.95\linewidth]{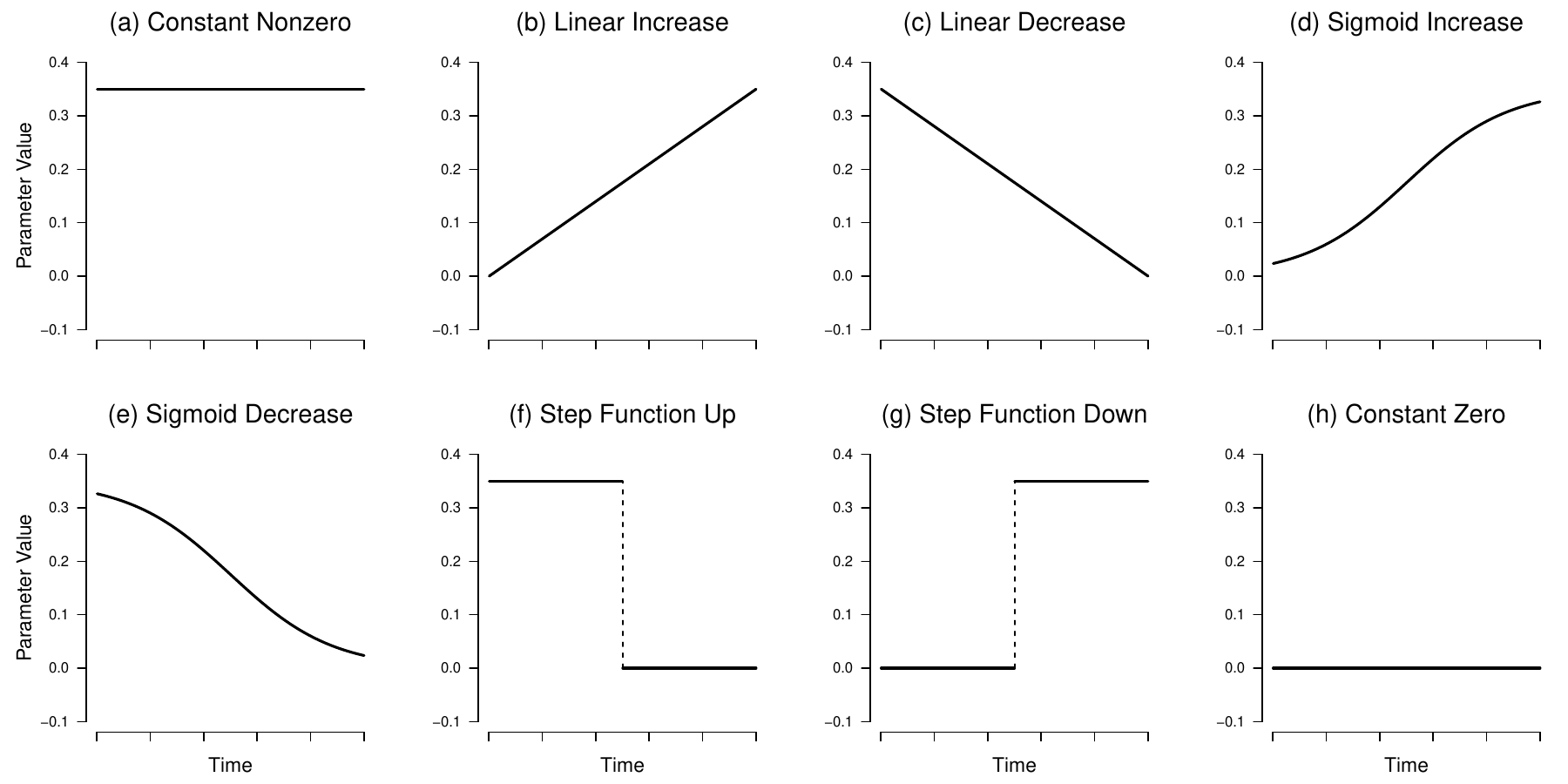}
	\caption{The eight types of time-varying parameters used in the simulation study: (a) constant nonzero, (b) linear increase, (c) linear decrease, (d) sigmoid increase, (e) sigmoid decrease, (f) step function up, (g) step function down and (h) constant zero.}
	\label{fig_data_gen_8types}
\end{figure}

If an edge is absent in the initial matrix, all entries of the parameter sequence are set to zero (panel (h) in Figure \ref{fig_data_gen_8types}). Note that only the time-varying parameter functions (a - e) and (h) in Figure \ref{fig_data_gen_8types} are smooth functions of time. Therefore, the two methods presented in this paper are only consistent estimators for those types of time-varying parameters. They cannot be consistent estimators for the step-functions (f) and (g), however, we included them to investigate how closely the methods studied in this paper can approximate the step function as a function of $n$.

The maximum parameter size of time-varying parameters is set to  $\theta = 0.35$ (see Figure \ref{fig_data_gen_8types}). The noise is drawn from a multivariate Gaussian with variances $\sigma^2 = \sqrt{0.10}$ and all covariances being equal to zero. Hence the signal/noise ratio used in our setup is $\mathrm{S/N} = \frac{0.35}{0.10} = 3.50$. All intercepts are set to zero and the covariances between the noise processes assigned to each variable are zero.

Using these time-varying VAR model, we generate 12 independent time series with lengths $n = \{ 20, 30, 36, 69, 103, 155, 234, 352, 530, 798, 1201, 1808 \}$. We chose these values because they cover the large majority of scenarios applied researchers typically encounter. Each of these time-varying models covers the full time period $[0,1]$ and is parameterized by a $p \times p \times n$ parameter array $B_{i, j, t}$. For example, the  $B_{1, 2, 310}$ indicates the cross-lagged effect from variable 2 on variable 1 at the 310th measurement point, which occurs then at time point $310/530 \approx 0.59$, if there are in total 530 measurements. Importantly, in this setting increasing $n$ does not mean that the time period between the first and the last measurement of the time series becomes larger. Instead, we mean by a larger $n$ that more evenly spaced measurements are available in the same time period. This means that the larger $n$, the smaller the time interval between two adjacent measurements. That is, the data density in the measured time period increases with $n$, which is required to consistently estimate time-varying parameters \citep{robinson1989nonparametric}. This makes sense intuitively: if the goal is to estimate the time-varying parameters of an individual in January, then one needs sufficient measurements in January, and it does not help to add additional measurements from February. 

We run 100 iterations of this design and report the mean absolute error over iterations. These mean errors serve as an approximation of the expected population level errors.

\subsubsection{Estimation}\label{sim_A_estimation}

To estimate time-varying VAR models via the GAM method we use the implementation in the R-package \textit{tvvarGAM} \citep{tvvarGAM} version 0.1.0, which is a wrapper around the \textit{mgcv} package (version 1.8-22). The tuning parameter of the spline method is the number of basis functions used in the GAM. Previous simulations have shown that $10$ basis functions give good estimates of time-varying parameters \citep{bringmann2018modeling}. To ensure that the model is identified, for a given number of basis functions $k$ and variables $p$, we require at least $n_\text{min} > k(p+1)$ observations. In our simulation, we used this constraint to select the maximum number of basis functions possible given $n$ and $p$, but we do not use less than 3 or more than 10 basis functions. That is, the selected number of basis functions $k_s$ is defined as 
	
\begin{equation}\label{knot_assignment}
k_s = \max \left  \{3,  \min \left  \{ \max \left \{ k ; k > \frac{n}{p+1} \right \}, 10 \right \} \right \}.
\end{equation}
	
If $k_s$ satisfies the above constraint, the time-varying VAR model can be estimated with the spline-based method. With this constraint the model cannot be estimated for $n = \{20, 30\}$. We therefore do not report results for GAM and GAM(st) for these sample sizes.

In principle it would be possible to combine $\ell_1$-regularization with the GAM-method, similarly as in the KS-method. However, an implementation of such a method is currently not available and we therefore cannot include it in our simulation.

We estimated the time-varying VAR model via the KS and KS(L1) methods using the R-package \emph{mgm} \citep{haslbeck2015mgm} version 1.2-2. As discussed in Section \ref{methods_ks}, these methods require the specification of a bandwidth parameter. Therefore, the first step of applying these methods is to select an appropriate bandwidth parameter by searching the candidate sequence $\bm{b} = \{0.01, 0.045, 0.08, 0.115, 0.185, 0.22, 0.225, 0.29, 0.325, 0.430, 0.465, 0.5 \}$. For $n \leq 69$ we omit the first 5 values in $\bm{b}$, and for $n > 69$ we omit the last 5 values. We did this to save computational cost because for small $n$, small $b$ are never selected, and analogously for large $n$, large $b$ values are never selected.  To select an appropriate bandwidth parameter we use a cross-validation-like scheme, which repeatedly divides the time series in a training and a test set, and in each repetition fits time-varying VAR models using the bandwidths in $b$, and evaluates the prediction error on the test set for each bandwidth. 
More concretely, we define a test set $S_{\mathrm{test}}$ by selecting $|S_{\mathrm{test}}| = \lceil (0.2n)^{2/3} \rceil$ time points stratified equally across the whole time series. Next, we estimate a time-varying VAR model for each variable $p$ at each time point in $S_{\mathrm{test}}$ and predict the $p$ values at that time point. After that we compute for each $b$ the $|S_{\mathrm{test}}| \times p$ absolute prediction errors and take the arithmetic mean. Next, we select the bandwidth $\hat{b}$ that minimizes this mean prediction error. Finally, we estimate the model on the full data using $\hat{b}$ and $\hat{\lambda}$ at 20 equally spaced time points, where we select an appropriate penalty parameter $\hat{\lambda}_i$ with 10-fold cross-validation for each of the $p$ variables \citep[for more details see][]{haslbeck2015mgm}.

We also investigate the performance of the kernel-smoothing method without $\ell_1$-regularization. We refer to this method as KS. In order to compare the $\ell_1$-regularized time-varying VAR estimator to a stationary  $\ell_1$-regularized VAR estimator, we also estimate the latter using the \emph{mgm} package. We call this estimator GLM(L1).

Both time-varying estimation methods are consistent if the following assumptions are met; (a) the data is generated by a time-varying VAR model as specified in equation (\ref{eq_VAR}), (b) all parameters are smooth functions of time, (c) with the eigenvalues of the VAR matrix being within the unit circle at all time points, (d) and the error covariance matrix is diagonal. We also fit a standard stationary VAR model using linear regression to get the unbiased stationary counter-part of the GAM methods. Specifically for the KS-method, it is additionally required that we consider small enough candidate bandwidth values. We do this by using the sequence $\bm{b}$ specified above.

\subsubsection{Results}\label{sim_A_results}

We first report the performance of the GLM, GLM(L1), KS, KS(L1), GAM and GAM(st) methods in estimating different time-varying parameters by evaluating the estimation error \textit{averaged across time}. Next, we zoom in on the performance \textit{across time}, for the constant and the linear increasing parameter function, and finally examine the performance in structure recovery of all methods.

\paragraph{Absolute Error Averaged over Time}\label{results_absolute}

\begin{figure}[H]		
		\includegraphics[width=.95\linewidth]{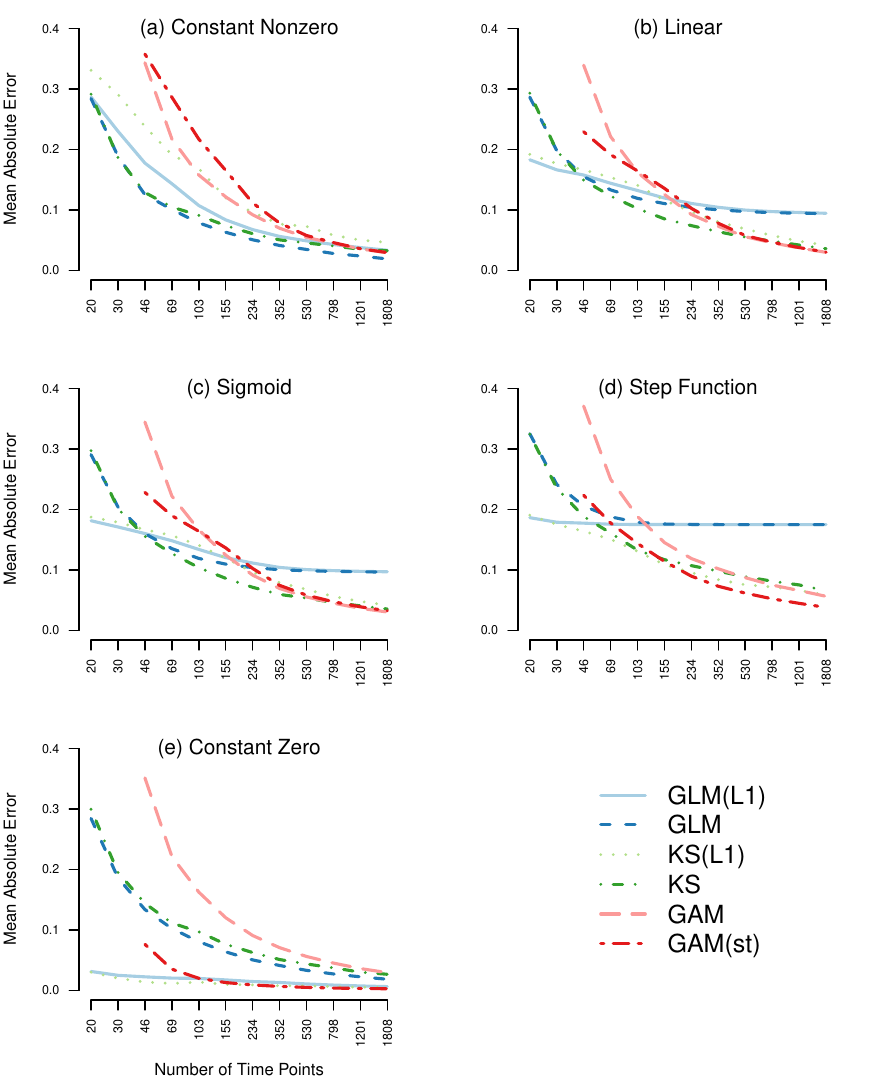}
		
		\caption{The five panels show the mean absolute estimation error averaged over the same type, time points, and iterations as a function of the number of observations $n$ on a log scale. We report the error of six estimation methods: stationary unregularized regression (blue), stationary $\ell_1$-regularized regression (light blue), time-varying  regression via kernel-smoothing (green), time-varying $\ell_1$-regularized regression via kernel-smoothing (light green), time-varying regression via GAM (pink), and time-varying regression via GAM with thresholding at 95\% CI (red). Some data points are missing because the respective models are not identified in that situation (see Section \ref{sim_A_estimation}).}
		\label{fig_simA}
	\end{figure}

Figure \ref{fig_simA} displays the absolute estimation error, averaged over time points, iterations, and time-varying parameter functions of the same type, as a function of sample size $n$. Since the linear increase/decrease, sigmoid increase/decrease, and step function increase/decrease are symmetric, we collapsed them into single categories to report estimation error. The absolute error on the y-axis can be be interpreted as follows: let's say we are in the scenario with $n = 155$ observations and estimate the constant function in Figure \ref{fig_simA} (a) with the stationary $\ell_1$-regularized regression GLM(L1). Then the \emph{expected} average (across the time series) error of the constant function is $\pm 0.09$.

Figure \ref{fig_simA} (a) shows that, for all methods, the absolute error in estimating the constant nonzero function is large for small $n$ and seems to converge to zero as $n$ increases. The GLM method has a lower estimation error than its $\ell_1$-regularized counterpart, GLM(L1). Similarly, the KS method outperforms the KS(L1) method. The stationary GLM method also outperforms all time-varying methods, which makes sense because the true parameter function is not time-varying. 

For the linearly increasing/decreasing time-varying parameter in Figure \ref{fig_simA} (b), the picture is more complex. For very small $n$ from 20 to 46 the regularized methods GLM(L1) and KS(L1) perform best. This makes sense because, for such small $n$, the estimates of all other methods suffer from huge variance. For sample sizes from 46 to 155 the unregularized methods perform better: now the bias of the regularized methods outweighs the reduction in variance. From sample sizes between 155 and 352 the time-varying methods start to outperform the two stationary methods. Interestingly, until around $n=530$ the KS methods outperforms all other time-varying methods. For $n>530$ all time-varying methods perform roughly equally. Overall, the error of all time-varying methods seem to converge to zero, as we would expect from a consistent estimator. The error of the stationary methods converges to $\approx 0.088$, which is the error resulting from approximating the time-varying function with the optimal constant function with value $\frac{0.35}{2}$. Since the sigmoid increase/decrease functions in panel (c) are very similar to the linear increase/decrease functions, we obtain qualitatively the same results as in the linear case.

In the case of the step function we again see a similar qualitative picture, however here the time-varying methods outperform the stationary methods already at a sample size of around $n=69$. The reason is that the step function is more time-varying in the sense that here the best constant function is a worse approximation than in the linear and the sigmoid case. Another difference is that the GAM(st) method seems to outperform all other methods by a small margin if the sample size is large.

Finally, the absolute error for estimating the constant zero function is lowest for the regularized methods and the thresholded GAM method. This is what one expect since these methods bias estimates towards zero, and the true parameter function is zero across the whole time period.

In Figure \ref{fig_simA} we reported the mean population errors of the six compared methods in various scenarios. These mean errors allow one to judge whether the \emph{expected} error of one method will be larger than the one of another method. However, it is also interesting to inspect the population sampling variance around these mean errors. This allows one to gauge with which probability one method will be better than another for a given sample. We show a version of Figure \ref{fig_simA} that includes the 25\% and 95\% quantiles of the absolute error in Appendix \ref{app_abserror_CIs}.

\paragraph{Absolute Error over Time for Constant and Linear Increasing Function}\label{results_overtime}

\begin{figure}[H]
	\centering
	\includegraphics[width=1\linewidth]{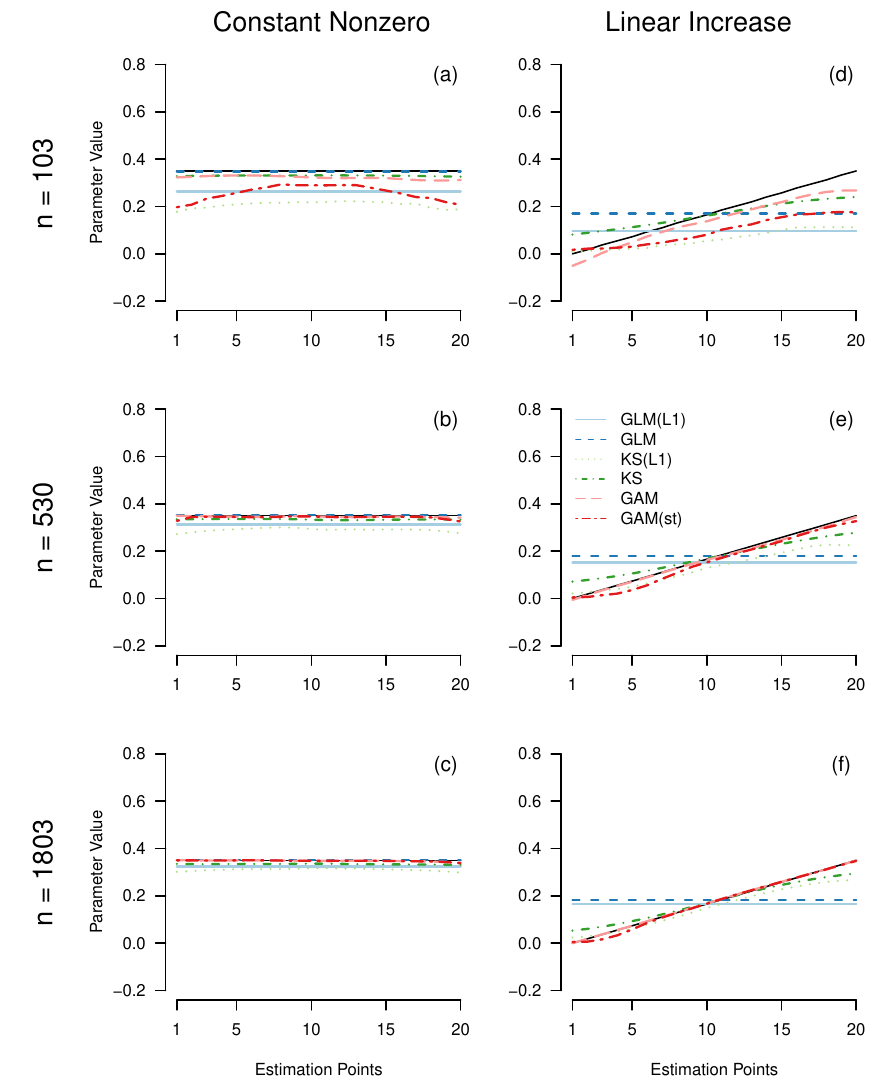}
	\caption{Mean and standard deviations of estimates for the constant parameter (left column), and the linear increasing parameter (right column), for $n = 103$ (top row),  $n = 530$ (second row) and $n = 1803$ (bottom row) averaged over iterations, separately for the five estimation methods: stationary $\ell_1$-regularized regression (red), unregularized regression (blue), time-varying $\ell_1$-regularized regression via kernel-smoothing (green), time-varying regression via GAM (pink), and time-varying regression via GAM with thresholding at 95\% CI (orange).}
	\label{fig_simB}
\end{figure}

To investigate the behavior of the different methods in estimating parameters across the time interval, Figure \ref{fig_simB} displays the mean absolute error for each estimation point (spanning the full period of the time series) for the constant nonzero function and the linear increasing function for $n = \{103, 530, 1803 \}$. Note that these results were already shown in aggregate form in Figure \ref{fig_simA}: for instance, the average (across time) of estimates of the stationary $\ell_1$-regularized method in Figure \ref{fig_simB} (a) corresponds to the single data point in Figure \ref{fig_simA} (a) of the same method at $n = 103$. 

Panel (a) of Figure \ref{fig_simB} shows the average parameter estimates of each method for the constant function with $n = 103$ observations. In line with the aggregate results in Figure \ref{fig_simA}, the stationary methods outperform the time-varying methods, and the unregularized methods outperform the regularized methods. We also see that the KS(L1) and the GAM(st) methods are biased downwards at the beginning and the end of the time series. The reason is that less data is available at these points, which results in stronger bias toward zero (KS(L1)) and more estimates being thresholded to zero. When increasing $n$, all methods become better at approximating the constant nonzero function. This is what we would expect from the results in Figure \ref{fig_simA}, which suggested that the absolute error of all methods converges to zero as $n$ grows.

In the case of the linear increase with $n = 103$ (d), we see that the time-varying methods follow the form of the true time-varying parameter, however, some deviations exists. With larger $n$, the time-varying methods recover the linearly increasing time-varying parameter with increasing accuracy. In contrast, the stationary methods converge to the best-fitting constant function. We also see that the average estimates of the regularized methods are closer to zero than the estimates of the unregularized methods. However, note that, similar to panel (e) in Figure \ref{fig_simA}, the regularized methods would perform better in recovering the constant zero function.

Here we only presented the mean estimates of each method, which displays the bias of the different methods as a function of sample size. However, it is equally important to consider the variance around estimates. We display this variance in Figure \ref{fig_simB_wCIs} in Appendix \ref{app_abserrorOT_CIs}. This figure shows that --- as expected --- the variance is very large for small $n$, but approaches 0 when $n$ becomes large.

\paragraph{Performance in Structure Recovery}\label{results_structure}

In some situations the main interest may be to recover the \emph{structure} of the VAR model, that is, we would like to know which parameters in the VAR parameter matrix are nonzero. We use two measures to quantify the performance of structure recovery. Sensitivity, the probability that a parameter that is nonzero in the true model is estimated to be nonzero; and precision, the probability that a nonzero estimate is nonzero in the true model. While higher values are better for both sensitivity and precision, different estimation algorithms typically offer different trade-offs between the two. Figure \ref{fig_simC} shows this trade-off for the five estimation methods:

\begin{figure}[H]
	
	\centering
	\includegraphics[width=.9\linewidth]{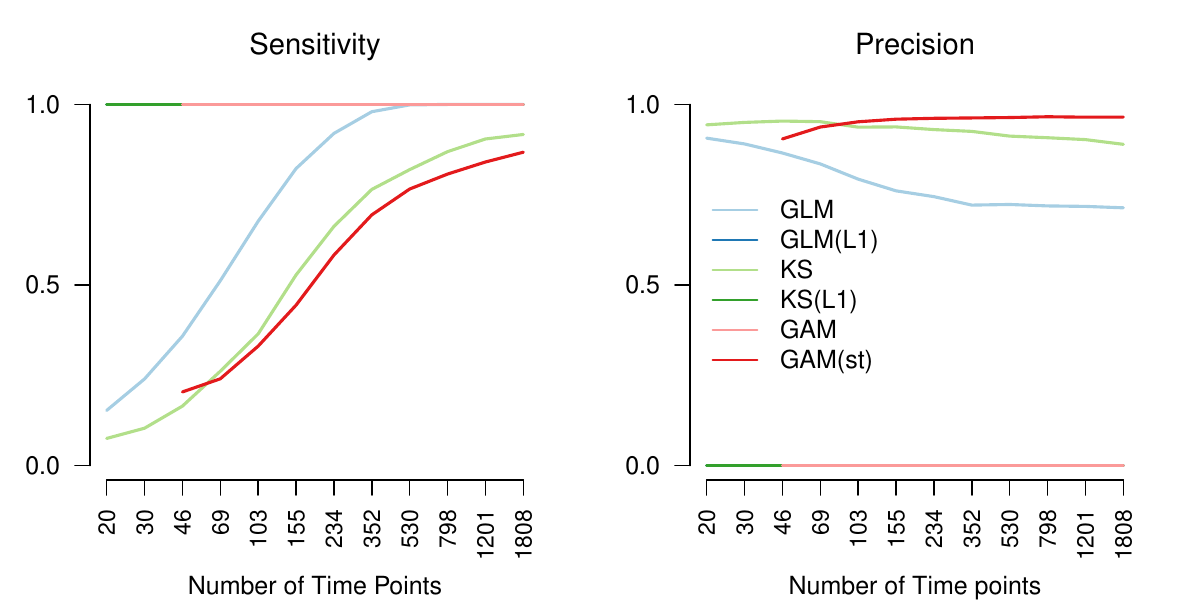}
	\caption{Sensitivity and precision for the five estimation methods across all edge-types for different variations of $n$. The lines for the unthresholded GAM(st) method and the stationary GAM method overlap completely, since they do not return estimates that are exactly zero. Some data points are missing because the respective models are not identified in that situation (see Section \ref{sim_A_estimation}).}
	\label{fig_simC}
\end{figure}

The unregularized stationary GLM method, the unregularized KS method, and the unthresholded time-varying GAM method have a sensitivity of 1 and a precision of 0 for all $n$. This is trivially the case because these methods return nonzero estimates with probability 1, which leads to a sensitivity of 1 and a precision of 0. Consequently, these methods are unsuitable for structure estimation. For all remaining methods, sensitivity seems to approach 1 when increasing $n$, while GLM(L1) has the highest sensitivity followed by KS(L1) and GAM(st). As expected, the precision of these methods is stacked up in reverse.

\paragraph{Computational Cost} In Appendix \ref{results_comptime} we report the computational cost of the time-varying methods. The main take away from these results is that computation time is not a major concern for typical psychological applications.

\subsubsection{Discussion}\label{sim_A_discussion}

The first simulation showed how the different methods perform in recovering a VAR model with $p=10$ variables based on a random graph, with linear, sigmoid, step and constant parameter functions, with sample sizes that cover most applications in psychology. The compared methods differ in the dimensions stationary vs. time-varying methods, unregularized vs. regularized methods, and GAM- vs. KS-based methods. Since all these dimensions interact with each other and with the type of time-varying parameter function they aim to recover, we discuss these interactions separately for each parameter function.

\paragraph{Constant Nonzero Function}
In the case of the constant nonzero function the stationary and unregularized GLM performed best, followed by the unregularized time-varying KS method. It makes sense that GLM performs best, because the true parameter function in this case is nonzero and constant across time. The KS method performs similarly especially for small $n$, because the bandwidth selection will select a very high bandwidth, which leads to a weighting that is almost equal for all time points, which leads to estimates that are very similar to the ones of the GLM method. The next best method is the stationary regularized GLM(L1) method. This is because the regularization decreases performance if the true parameter function is nonzero, however, it uses the correct assumption that the true parameter function is constant across time. Since the ability to estimate time-varying parameters is no advantage when estimating the constant nonzero function, the KS(L1) method performs worse than the GLM(L1) method. Interestingly, the unregularized GAM function performs much worse than the unregularized KS method. The significance-thresholded GAM(st) method performs worse than the GAM method, because if the true parameter function is nonzero, thresholding it to zero can only increase estimation error.

\paragraph{Linear and Sigmoid Functions}
The results for the linear increasing/decreasing function are similar to the constant nonzero function, except that that all time-varying methods have a lower absolute error than the stationary methods from $n > 234$. The KS method is already better from $n > 46$. A difference to the constant nonzero function is that the two regularized methods GLM(L1) and KS(L1) perform best if the sample size is very small ($n < 46$). A possible explanation for this difference is that the bias toward zero of the regularization is less disadvantageous for the linear increasing/decreasing functions, because its parameter values are on average only half as large as for the constant nonzero function. Within time-varying functions, the KS method performs better than the KS(L1) methods, which makes sense because the true parameter function is nonzero. For the same reason, the GLM method outperforms the GAM(st) method. The KS methods perform better than the GAM methods for sample sizes up to $n=530$. The reason is that the estimates of the GAM methods have a larger sampling variance (see Figure \ref{fig_simA_app} in Appendix \ref{app_abserror_CIs}). The errors in estimating the sigmoid function are very similar to the linear increasing/decreasing functions, since their functional forms are very similar. 

\paragraph{Step Function}
The errors in estimating the step function are again similar to the linear and the sigmoid case, except for two differences: first, the time-varying methods become better than the stationary methods already between $n=46$ and $n=69$. And second, the regularized KS(L1) performs better than KS, and the thresholded GAM(st) method performs better than the GAM method. The reason is that in half of the time series the parameter value is zero, which can be recovered exactly with the KS(L1) and the GAM(st) methods. This advantage seem to outweigh the bias these methods have in the other half of the time series in which the parameter function is nonzero.

\paragraph{Constant Zero Function}
In the case of the constant zero function the errors are roughly stacked up the reverse order as in the constant nonzero function. The regularized GLM(L1) and KS(L1) do best, followed by the thresholded GAM(ks) method. Among the unregularized methods the GLM and KS methods perform quite similarly, with the GLM method being slightly better, because the true parameter function is constant. Interestingly, the GAM method performs far worse, which is again due to its high variance (see Figure \ref{fig_simA_app} in Appendix \ref{app_abserror_CIs}).

\paragraph{Summary}
We saw that stationary methods outperform time-varying methods when the true parameter function is a constant, and time-varying methods out-perform stationary methods if the true parameter function is time-varying, and if the sample size is large enough. The sample size at which the time-varying methods become better depends on how time-varying the true parameter is: the more time-varying it is, the smaller the sample size $n$ at which time-varying methods become better than stationary ones. Within time-varying methods, the KS methods outperformed the GAM methods for smaller sample sizes, while the GAM based methods became better with very large sample sizes ($n > 530$). 

Finally, we saw that regularized methods perform better if the true parameter function is zero, while unregularized methods perform better if the true parameter function is nonzero, as expected. In order to choose between regularized and unregularized methods, one therefore needs to judge how many of the parameters in the true time-varying VAR model are nonzero. Given the expected sparsity of the true VAR model, one could compute a weighted average of the errors shown in this section in order to determine which method has the lowest overall error. However, to evaluate the performance of the different methods for different levels of sparsity more directly, we performed a second simulation study in which we vary the sparsity of the VAR model.

\subsection{Simulation B: Varying Sparsity}\label{sim_B_UT}

In this simulation we evaluate the absolute estimation error of all methods for the different parameter functions and for the combined time-varying VAR model, as a function of sparsity. Specifically, we evaluate the estimation error of recovering the time-varying predictors of a given variable in the VAR model, depending on how many predictors are nonzero. From a network perspective the number of predictors on a given node is equal to its indegree. We will vary the indegree from 1 to 20. The average indegree in Simulation A was  $1 + 9 \times P(\text{edge}) = 2.61$.

\subsubsection{Data Generation}

We vary sparsity by specifying the structure of the initial VAR matrix to be upper-triangular. We show the structure of such a matrix, and the corresponding directed network in Figure \ref{fig_sim_UT}:

\begin{figure}[H]
			\begin{minipage}[b]{0.55\linewidth}

								$$
								\scalemath{0.9}{
									\bordermatrix{   &X_{1, t-1}&X_{2, t-1}&X_{3, t-1}&X_{4, t-1} &X_{5, t-1} &X_{6, t-1}\cr
										X_{1, t}     & 1  &  1    & 1       & 1 & 1 & 1   \cr
										X_{2, t}     & 0    &  1  & 1    	  & 1  & 1 & 1  \cr
										X_{3, t}     & 0    &  0   	  & 1     & 1 & 1 & 1 \cr
										X_{4, t}     & 0    &  0   	  & 0     & 1 & 1 & 1\cr
										X_{5, t}     & 0    &  0   	  & 0     & 0 & 1 & 1\cr
										X_{6, t}     & 0    &  0  & 0    & 0 &0  & 1}
								}
								$$
								\vspace{.8cm}

			\end{minipage}
			\begin{minipage}[b]{0.45\linewidth}
				
				\centering
				\vspace{.5cm}
				\includegraphics[width=.7\linewidth]{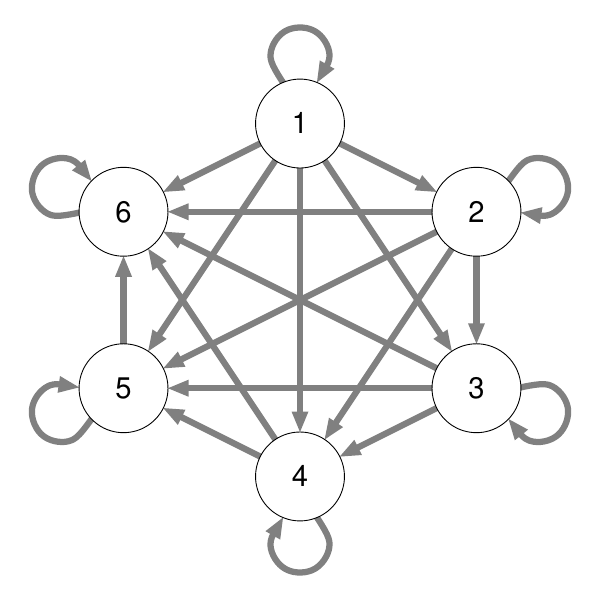}
				\vspace{0.2cm}

\end{minipage}

	\caption{
		Left: the upper-diagonal pattern of nonzero parameters used in the time-varying VAR model in the second simulation, here shown for six variables. The row sums are equal to the indegree of the respective nodes, which results in a frequency of one for each indegree value. Right: visualization of the upper-diagonal pattern as a directed graph. The graph used in the simulation has the same structure but is comprised of 20 nodes.
 		}
	\label{fig_sim_UT}
\end{figure}

In such a model, the first variable has one predictor (itself at $t-1$), the second variables has two predictors (itself and variable 1 at $t-1$), the third variable has three predictors, etc. and the last variable has $p$ predictors. As defined in Section \ref{sec_estimationmethods}, the number of nonzero predictor variables (or the indegree from a network perspective) is a local (i.e. for some variable X) measure of sparsity. In the simulation we use the same initial VAR matrix, except that we use a VAR model with $p=20$ variables. All additional steps of the data generation (see Section \ref{sim_A_datagen}, and the specification of the estimation methods (Section \ref{sim_A_estimation}) are the same as in Simulation A.

\subsubsection{Results}

\begin{figure}[H]
	\centering		
	\includegraphics[width=1\linewidth]{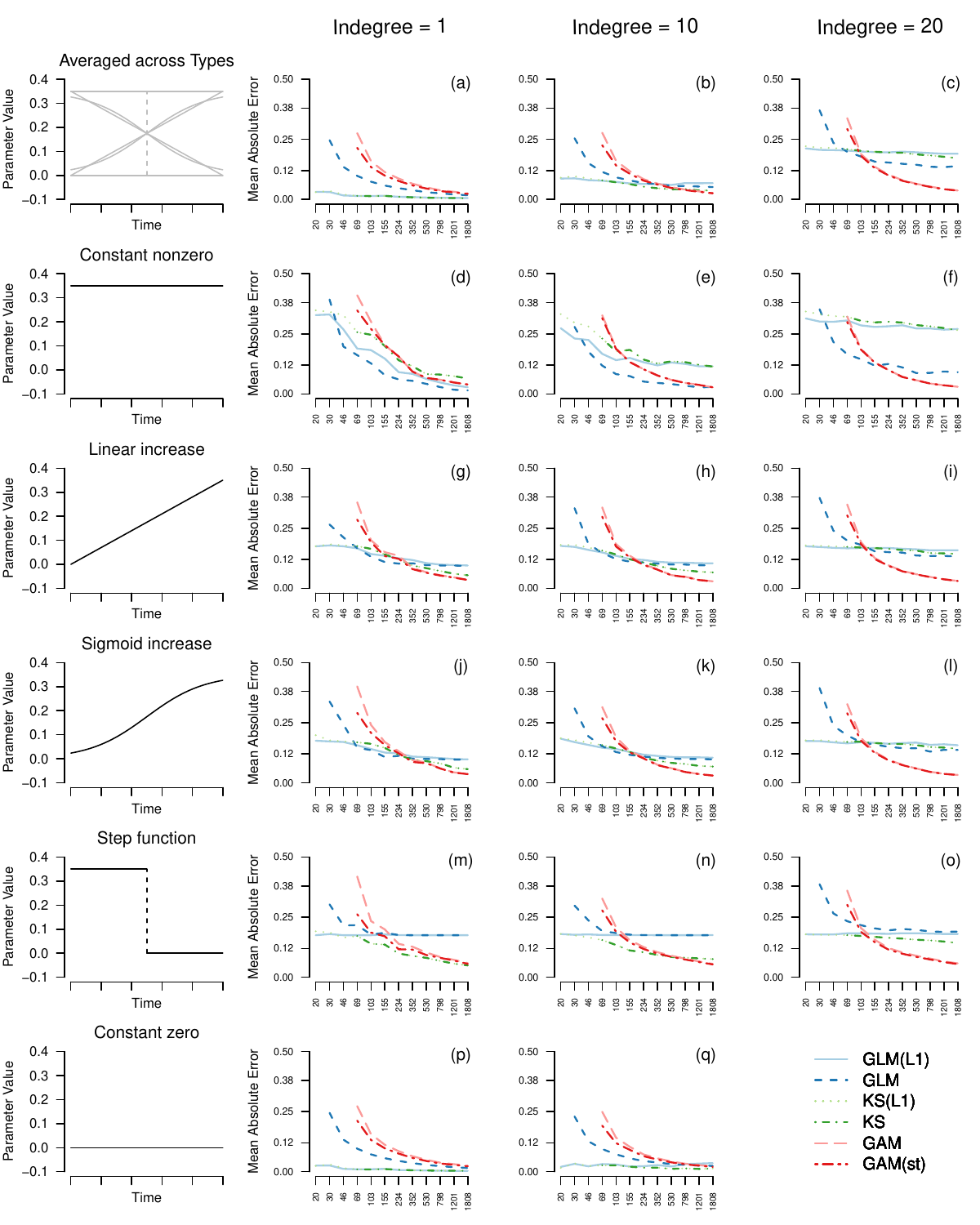}
	\caption{The mean average error for estimates of the upper-triangular model for all five estimation methods for the same sequence of numbers of time points $n$ as in the first simulation. The results are conditioned on three different indegrees (1, 10, 20)  and shown averaged across (a - c) and separately for the time-varying parameter types (d - q).}
	\label{fig_SIM_UT_results}
\end{figure}

Figure \ref{fig_SIM_UT_results} displays the mean absolute error separately for the five different time-varying parameter functions and for indegrees 1, 10, 20. Similarly to Simulation A, we collapsed symmetric increasing and decreasing functions into single categories and report their average performance. The first row of Figure \ref{fig_SIM_UT_results} shows the performance averaged over time points and types of time-varying parameters for indegree 1, 10 and 20. The most obvious result is that all methods become worse when increasing the indegree. This is what one would expect since more parameters are nonzero and more predictors are correlated. In addition, there are several interactions between indegree and estimation methods. First, the regularized methods perform best when indegree is low, and worst when indegree is high. This makes sense: the bias toward zero of the regularization is more beneficial if almost all parameter functions are zero. However, if most parameter functions are nonzero, a bias toward zero leads to high estimation error. Second, we see that the drop in performance is lower for the GAM based methods compared to the KS based methods. The combined results in the first row are the weighted average of the remaining rows. The estimation errors for the time-varying functions show a similar pattern as in Figure \ref{fig_simA} of Simulation A, except that the GAM methods perform better for indegree values 10 and 20.

\subsubsection{Discussion}

The results of Simulation B depicted the relative performance of all methods as a function of sparsity, which we analyzed locally as indegree. As expected, regularized methods perform better when indegree is low and worse if indegree is high. Interestingly, among the time-varying methods, the GAM based methods perform better than the KS based methods when indegree is high.

\subsection{Overall Discussion of Simulation Results}\label{sim_OverallDiscussion}

Here we discuss the overall strengths and weaknesses of all considered methods in light of the results of both simulations.

\paragraph{Stationary vs. Time-Varying Methods}
We saw that stationary methods outperform time-varying methods if the true parameter function is constant, as one would expect. If the parameter function is time-varying, then the stationary methods are better for very small sample sizes, but for larger sample sizes, the time-varying methods become better. The exact sample size $n$ at which time-varying methods start to perform better depends on how strongly the true parameters vary with time: the stronger the variation, the smaller the $n$. For the choice of true parameter functions in our simulations, we found that the best time-varying method outperformed the stationary methods at already $n > 46$.

\paragraph{Unregularized vs. Regularized Methods} The results in both simulations showed that if most true parameter functions are zero (high sparsity), regularized methods and the thresholded GAM(st) method performed better compared to their unregularized/unthresholded counterparts. On the other hand, if most true parameter functions are nonzero (low sparsity), the unregularized/unthresholded functions perform better. In Simulation B we specifically mapped out the performance of methods as a function of sparsity and found that unregularized methods are better at an indegree of 10 or larger.

\paragraph{Kernel-smoothing vs. GAM Methods}
If sparsity is high, that is, if most parameter functions are zero, the KS based methods outperformed the GAM based methods for most sample size regimes. Only if the sample size is very large the GAM based methods showed a performance that is equal or slightly better than the KS based methods. However, if sparsity is low, the GAM based methods outperformed the KS based methods.\\

Accordingly, applied researchers should choose the KS based methods when they expect the time-varying VAR model to be relatively sparse and if they only have a moderate sample size ($n < 200 \text{to} 300$). If one expects that only few parameter functions are nonzero, the KS based method should be combined with regularization. If one expects the parameter functions of the time-varying VAR model to be largely nonzero, and if one has a large sample size, the GAM based methods are likely to perform better.

\paragraph{Limitations}
Several limitations of the simulation studies require discussion. First, the signal to noise ratio $ \mathrm{S/N} = \frac{\theta}{\sigma} = 3.5$ in parameter values could be larger or smaller in a given application and the performance results would accordingly be better or worse. Similarly, the signal to noise ratio would be smaller if we increased the number of variables $p$, because more parameters have to be estimated. However, note that $\mathrm{S/N}$ is also a function of $n$. Hence if we assume a lower $\mathrm{S/N}$ this simply means that we need more observations to obtain the same performance, while all qualitative relationships between time-varying parameters, structure in the VAR model and estimators remain the same.

Second, the time-varying parameters could be more time-varying. For example, we could have chosen functions that go up and down multiple times instead of being monotone increasing/decreasing. However, for estimation purposes, the extent to which a function is time-varying is determined by how much it varies over a specified time period \emph{relative} to how many observation are available in the time period. Thus the $n$-variations can also be seen as a variation of the extent to which parameters are varying over time: From this perspective, the time-varying parameter functions with $n = 20$ are very much varying over time, while the parameter functions with $n = 1808$ are hardly varying over time. Since we chose $n$-variations stretching from unacceptable performance ($n = 20$) to very high performance ($n = 1808$), we simultaneously varied the extent to which parameters are time-varying.

Third, we only investigated time-varying VAR models with $p=10$ variables and a single lag. In terms of the performance in estimating (time-varying) VAR parameters, adding more variables or lags boils down to increasing the indegree of a VAR model with a single lag and fixed $p$. In general, the larger the indegree and the higher the correlations between the predictors, the harder it is to estimate the parameters associated with a variable. Part of the motivation for Simulation B in Section \ref{sim_B_UT} was to address this limitation.

Finally, we would like to stress that all statements with respect to sample size refer to the effective sample size available to estimate the VAR model. We mention this because the effective sample size that is used to estimate a VAR model is often considerably lower than the number of measurement points in an ESM study. This is both because of planned (e.g., at the day/night shift) and unplanned missing values. For example, if an ESM study has five measurements a day with a measurement interval of 3h and the fourth measurement is missing, then the effective sample size is only three, because only for three time points (2, 3, and 4) a measurement 3h before is available.


\section{Estimating time-varying VAR model on Mood Time Series}\label{sec_tutorial}

In this section we provide a step-by-step tutorial on how to estimate a time-varying VAR model on a mood time series using the KS(L1) method. In addition, we show how to compute time-varying prediction errors for all nodes, how to assess the reliability of all estimates, and how to visualize some aspects of the estimated time-varying VAR model. Finally, we briefly discuss how to select between stationary and time-varying models. All analyses are performed using the R-package \textit{mgm} (version 1.2-8) \citep{haslbeck2015mgm} and R-version 3.6.0, and the code below can also be found as an R-file on Github: \url{https://github.com/jmbh/tvvar_paper}. In Appendix \ref{sec_tut_gam} we show how to fit the same model with the GAM(st) method using the R-package \emph{tvvarGAM}.

\subsection{Data}

We illustrate how to fit a time-varying VAR model on a symptom time series with 12 variables related to mood measured on 1476 time points during 238 consecutive days from an individual diagnosed with major depression \citep{wichers2016critical}. The measurements were taken at 10 pseudo-randomized time intervals with average length of 90 minutes between 07:30 and 22:30. During the measured time period, a double-blind medication dose reduction was carried out, consisting of a baseline period, the dose reduction, and two post assessment periods (See Figure \ref{fig_app_mgm}, the points on the time line correspond to the two dose reductions). For a detailed description of this data set see \cite{kossakowski2017data}.

\subsection{Load R-packages and Dataset}\label{tutorial_mgm_data}

The above described symptom dataset automatically available when loading the R-package \textit{mgm}. After loading the package, we subset the 12 mood variables contained in this dataset:

\begin{Verbatim}[commandchars=\\\{\}, fontsize=\small]
library(mgm) # Version 1.2-8

mood_data <- as.matrix(symptom_data$data[, 1:12]) # Subset variables
mood_labels <- symptom_data$colnames[1:12] # Subset variable labels
colnames(mood_data) <- mood_labels
time_data <- symptom_data$data_time
\end{Verbatim}

The object \verb|mood_data| is a $1476 \times 12$ matrix with measurements of 12 mood variables:

\begin{Verbatim}[commandchars=\\\{\}, fontsize=\small]
> dim(mood_data)
[1] 1476   12

> head(mood_data[,1:7])
     Relaxed Down Irritated Satisfied Lonely Anxious Enthusiastic
[1,]       5   -1         1         5     -1      -1            4
[2,]       4    0         3         3      0       0            3
[3,]       4    0         2         3      0       0            4
[4,]       4    0         1         4      0       0            4
[5,]       4    0         2         4      0       0            4
[6,]       5    0         1         4      0       0            3
\end{Verbatim}

The matrix \verb|time_data| contains information about the time stamps of each measurement. This information is needed for the data preprocessing in the next section.

\begin{Verbatim}[commandchars=\\\{\}, fontsize=\small]
> head(time_data)
  date     dayno beepno beeptime resptime_s resptime_e   time_norm
1 13/08/12   226      1    08:58   08:58:56   09:00:15 0.000000000
2 14/08/12   227      5    14:32   14:32:09   14:33:25 0.005164874
3 14/08/12   227      6    16:17   16:17:13   16:23:16 0.005470574
4 14/08/12   227      8    18:04   18:04:10   18:06:29 0.005782097
5 14/08/12   227      9    20:57   20:58:23   21:00:18 0.006285774
6 14/08/12   227     10    21:54   21:54:15   21:56:05 0.006451726
\end{Verbatim}

For a sizable number of measurement points the individual did not provide a response. The \textit{mgm} package takes care of this automatically, by only using those time points to estimate a VAR(1) model for which a measurement at the previous time point is available.

Some of the variables in this data set are highly skewed, which can lead to unreliable parameter estimates. Here we deal with this issue by computing bootstrapped confidence intervals (KS method) and credible intervals (GAM method), to judge how reliable the estimates are. Since the focus in this tutorial is on estimating time-varying VAR models, we do not investigate the skewness of variables in detail. However, in practice the marginal distributions should always be inspected before fitting a (time-varying) VAR model. A possible remedy for skewed variables is to transform them, typically by taking a root, the log, or transformations such as the nonparanormal transform \citep{liu2009nonparanormal}. A disadvantage of this approach is that the parameters are more difficult to interpret. For example, if applying the log-transform to $X$, then the cross-lagged effect $\beta_{X, Y}$ of $Y$ on $X$ is interpreted as ``When increasing $Y$ at $t-1$ by 1 unit, the log of $X$ at $t$ increases by $\beta{X, Y}$, when keeping all other variables at $t-1$ constant''.

\subsection{Estimating Time-Varying VAR Model}\label{tut_KS_estimation}

Here we describe how to use the function \verb|tvmvar()| of the \textit{mgm} package to estimate a time-varying VAR model. A more detailed description of this function can be found in the help file \verb|?tvmvar|. After providing the data via the \verb|data| argument, we specify the type and levels of each variable. The latter is necessary because \textit{mgm} allows one to estimate models including different types of variables. In the present case we only have continuous variables modeled as conditional Gaussian distributions, and we therefore specify \verb|type = rep("g", 12)|. By convention the number of levels for continuous variables is specified as one \verb|level = rep(1, 12)|. 

Via the argument \verb|estpoints| we specify that we would like to have 20 estimation points that are equally spaced across the time series (for details see  \verb|?tvmvar|). The number of estimation points can be chosen arbitrarily large, however at some point adding more estimation points becomes useless because adjacent estimation points become indistinguishable. Via the argument \verb|timepoints| we provide a vector containing the time point of each measurement. The time points are used to distribute the estimation points correctly on the time interval. If no \verb|timepoints| argument is provided, the function assumes that all measurement points are equidistant. See Section 2.5 in \cite{haslbeck2015mgm} for a more detailed explanation how the time points are used in \textit{mgm} and an illustration of the problems following from incorrectly assuming equidistant measurement points.

Next, we specify the bandwidth parameter $b$, which determines how many observations close to an estimation point are used to estimate the model at that point. Here we select $b = 0.34$, which we obtained by searching a candidate sequence of bandwidth parameters, and selected the value that minimized the out-of-bag cross-validation error. The latter is implemented in the function \verb|bwSelect()| (for details on this time-stratified cross-validation scheme see Section \ref{sim_A_estimation}). Since \verb|bwSelect()| repeatedly fits time-varying VAR models with different bandwidth parameters, the specification of \verb|bwSelect()| and the estimation function \verb|tvmvar| are very similar. We therefore refer the reader for the code to specify \verb|bwSelect()|  to Appendix \ref{app_bandwidth}.

After that we provide the number of the notification on a given day and the number of the day itself via the arguments \verb|beepvar| and \verb|dayvar|, respectively. This information is used to exclude cases from the analysis which do not have sufficient previous measurements to fit the specified VAR model. This can be both due to randomly missing data, or because of missingness by design. In the present dataset we have both: within a given day the individual did not always answer at all 10 times. And by design, there is a break between day and night. When not considering the correct successiveness the estimated parameters do not only reflect effects from $t_{t-1}$ on $t$ but also effects over (possibly) many other time-lags (for instance 10h over night instead of the intended 1h30).

Via the argument \verb|lags = 1| we specify to fit a first order VAR model and specify with the argument \verb| lambdaSel = "CV"|  to select the penalty parameters $\lambda$ with cross-validation. Finally, with the argument \verb|scale = TRUE| we specify that all variables should be scaled to mean zero and standard deviation 1 before the model is fit. This is recommended when using $\ell_1$-regularization, because otherwise the strength of the penalization of a parameter depends on the variance of the predictor variable \citep[see][p.9]{hastie2015statistical}. Since the cross-validation scheme uses random draws to define the folds, we set a seed to ensure reproducibility. 

\begin{Verbatim}[commandchars=\\\{\}, fontsize=\small]
set.seed(1)
tvvar_obj <- tvmvar(data = mood_data,
                    type = rep("g", 12),
                    level = rep(1, 12), 
                    lambdaSel = "CV",
                    timepoints = time_data$time_norm, 
                    estpoints = seq(0, 1, length = 20), 
                    bandwidth = 0.34,
                    lags = 1,
                    beepvar = time_data$beepno,
                    dayvar = time_data$dayno,
                    scale = TRUE)
\end{Verbatim}

Before looking at the results we check how many of the 1476 time points were used for estimation, which is shown in the summary that is printed when calling the output object in the console:

\begin{Verbatim}[commandchars=\\\{\}, fontsize=\small]
> tvvar_obj
mgm fit-object 

Model class:  Time-varying mixed Vector Autoregressive (tv-mVAR) model 
Lags:  1 
Rows included in VAR design matrix:  876 / 1475 ( 59.39 %) 
Nodes:  12 
Estimation points:  20
\end{Verbatim}

This means that the VAR design matrix that is used for estimation has 876 rows. One of the removed time points is the first time point, since it does not have a previous time point. Other time points were excluded because of (a) missing measurements during the day or (b) the day-night break. As an example, from the six rows of the time stamps shown above, we could use three time points, since a measurement at the previous time point is available.

The absolute values of the estimated VAR coefficients are stored in the object \verb|tvvar_obj$wadj|, which is an array of dimensions $p \times p \times \text{lags} \times \text{estpoints}$, lags is the number of lags, and estpoints is the number of estimation points. For example, the array entry \verb|tvvar_obj$wadj[1, 3, 1, 9]| returns the cross-lagged effect of variable 3 on variable 1 with the first specified lag size (here 1) at estimation point 9. Due to the large number of estimated parameters, we do not show this object here but instead visualize some aspect of it in Figure \ref{fig_app_mgm}. The signs of all parameters are stored separately in \verb|tvvar_obj$signs|, because signs may not be defined in the presence of categorical variables (which is not the case here). The intercepts are stored in \verb|tvvar_obj$intercepts|.

\subsection{Assessing Reliability of Parameter Estimates}

To judge the reliability of parameter estimates, we approximate the sampling distribution of all parameters using the nonparametric block bootstrap. The function \verb|resample()| implements this bootstrap scheme and returns the sampling distribution and a selection of its quantiles of each parameter. First we provide the model object \verb|object = tvvar_obj| and the data \verb|data = mood_data|.  \verb|resample()| then fits the model specified as in \verb|tvvar_obj| on 50 (\verb|nB = 50|) different block bootstrap samples, where we specify the number of blocks via  \verb|blocks|. The argument \verb|seeds| provides a random seed for each bootstrap sample and \verb|quantiles| specified the quantiles shown in the output.

\begin{Verbatim}[commandchars=\\\{\}, fontsize=\small]
res_obj <- resample(object = tvvar_obj, 
                    data = mood_data, 
                    nB = 50, 
                    blocks = 10,
                    seeds = 1:50, 
                    quantiles = c(.05, .95))
\end{Verbatim}

The $p \times p \times \text{lags} \times \text{estpoints} \times nB$ array \verb|res_obj$bootParameters| contains the sampling distribution of each parameter. For instance, the array entry \verb|res_obj$bootParameters[1, 3, 1, 9, ]| contains the sampling distribution of the cross-lagged effect of variable 3 on variable 1 with the first specified lag size (here 1) at time point 9. Due to its size, we do not show this object here but show the 5\% and 95\% quantiles of the empirical sampling distribution of three time-varying parameters in Figure \ref{fig_app_mgm}. Also note that the resampling procedure is computationally expensive. With 50 bootstrap samples as specified above, the \verb|resample()| runs approximately 10 minutes.

It is important to keep in mind that the quantiles of these bootstrapped sampling distributions are not confidence intervals around the true parameter. The reason is that the $\ell_1$-penalty biases all estimates and hence the whole sampling distribution towards zero which implies that the latter is not centered on the true parameter value.

\subsection{Computing Time-Varying Prediction Error}

Here we show how to compute time-varying nodewise prediction errors. Nodewise prediction errors indicate how well the model fits the data on an absolute scale and is therefore useful to judge the practical relevance of (parts of) a VAR model. See \cite{haslbeck2018well} for a detailed description of nodewise prediction error (or predictability) in the context of network models and \cite{haslbeck2017predictable} for an analysis of predictability in 18 datasets in the field of psychopathology.

The function \verb|predict()| computes predictions and prediction errors from a given \textit{mgm} model object. We first provide the model object \verb|object = tvvar_obj| and the data \verb|data = mood_data|. We then specify the desired types of prediction, here \verb|R2| for the proportion of explained variance and \verb|RMSE| for the Root Mean Squared Error. \verb|tvMethod = "weighted"| specifies how to combine all time-varying models to arrive at a single prediction for each variable across the whole time series (for details see  \verb|?predict|). Finally, we provide \verb|consec = time_data$beepno| for the same reasons as above.

\begin{Verbatim}[commandchars=\\\{\}, fontsize=\small]
pred_obj <- predict(object = tvvar_obj, 
                    data = mood_data, 
                    errorCon = c("R2", "RMSE"),
                    tvMethod = "weighted", 
                    consec = time_data$beepno)
\end{Verbatim}

The predictions are stored in \verb|pred_obj$predicted| and the error of the predictions of all time-varying models combined are in \verb|pred_obj$errors|:

\begin{Verbatim}[commandchars=\\\{\}, fontsize=\small]
> pred_obj$errors
       Variable Error.RMSE Error.R2
1       Relaxed      0.939    0.155
2          Down      0.825    0.297
3     Irritated      0.942    0.119
4     Satisfied      0.879    0.201
5        Lonely      0.921    0.182
6       Anxious      0.950    0.086
7  Enthusiastic      0.922    0.169
8    Suspicious      0.818    0.247
9      Cheerful      0.889    0.200
10       Guilty      0.928    0.175
11        Doubt      0.871    0.268
12       Strong      0.896    0.195
\end{Verbatim}

The prediction errors of each time-varying model separately are stored in \verb|pred_obj$tverrors|. Note that here we weight the errors using the same weight vector as used for estimation (see Section \ref{methods_ks}). For details see \verb|?predict.mgm|. In the following section we visualize the time-varying nodewise estimation error for a subset of estimation points.

\subsection{Visualizing Time-Varying VAR model}

Figure \ref{fig_app_mgm} visualizes a part of the time-varying VAR parameters estimated above: 

\begin{figure}[H]
	\includegraphics[width=1\linewidth]{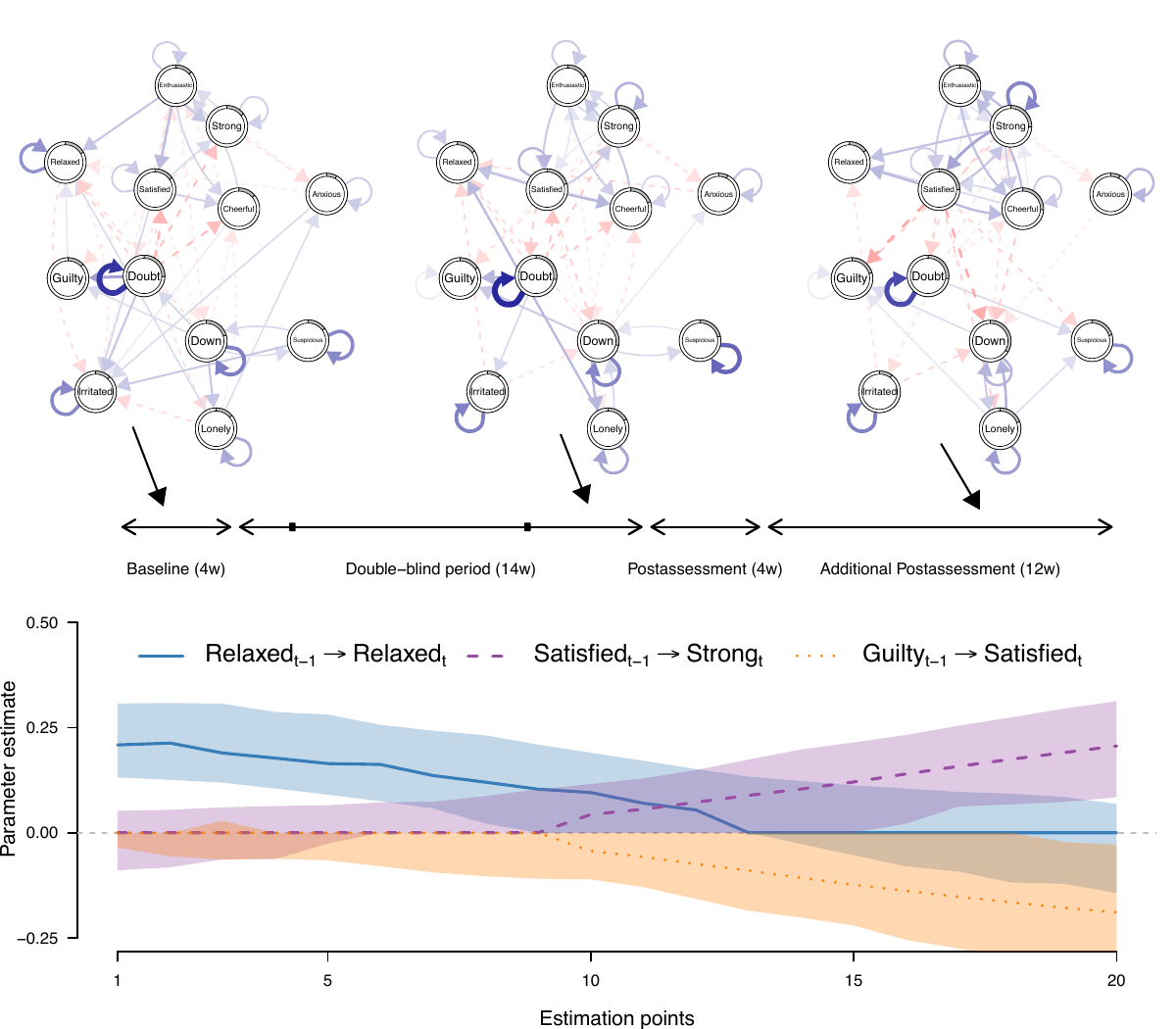}
	\caption{Top row: visualization of VAR(1) models at estimation points 2, 10 and 18. Blue solid arrows indicate positive relationships, red dashed arrows indicate negative relationships, and the width of the arrows is proportional to the absolute value of the corresponding parameter. The self-loops indicate autocorrelations. The colored parts of the ring around each node represents the respective within sample proportion of explained variance ($R^2$). Bottom row: three parameters plotted as a function of time; the points are the point estimate obtained from the full dataset, the shaded areas indicate the 5\% and 95\% quantiles of the bootstrapped sampling distribution at each estimation point. }
	\label{fig_app_mgm}
\end{figure}

The top row shows visualizations of the VAR parameters for the estimation points 2, 10 and 18. Blue solid arrows indicate positive relationships, red dashed arrows indicate negative relationships. The width of the arrows is proportional to the absolute value of the corresponding parameter. The grey part of the ring around each node indicates the proportion of explained variance of each variables by all other variables in the model. Comparing the VAR estimates across the three shown estimation points reveals that some parameters are strongly time-varying. For example, there is an autocorrelation effect of Relaxed at estimation point 2, which becomes smaller at estimation point 10 and vanishes at estimation point 18. On the other hand, the cross lagged effects $\mathrm{Satisfied}_{t-1} \rightarrow \mathrm{Strong}_{t}$ and $\mathrm{Satisfied}_{t-1} \rightarrow \mathrm{Strong}_{t}$ are equal to zero at estimation point 2 and become larger in estimation point 10 and 18. To better evaluate the time-varying nature of those three parameters we plot them as a line graph in the lower panel of Figure \ref{fig_app_mgm}. Aligning time-varying parameter functions with additional information available about an individual may allow one to explain the changes in parameters. For example, we see that the three time-varying parameters in the lower panel show their largest change after the second reduction of the antidepressant medication. This suggests that the medication reduction could be part of the explanation for this change in parameters. Next to individual interaction parameters, possible analyses can also focus on the changes in intercepts or aggregates of several parameters. For example, one could investigate how the density of the entire or parts of the VAR model changes across time. The code to fully reproduce Figure \ref{fig_app_mgm} is not shown here due to its length, but can be obtained from Github (\url{https://github.com/jmbh/tvvar_paper}).

\subsection{Selecting between Stationary and Time-varying Models}

While model selection between stationary and time-varying models is not the topic of this paper and requires a separate treatment to be addressed adequately, we briefly comment on this issue in relation to the methods presented here. One possible way to select between a stationary and a time-varying (VAR) model is to divide the time series into a training and test set. Then one can fit each model on the training set and evaluate on the test set which model has the lower prediction error. In fact, this is the procedure that is implemented in the function \verb|bwSelect()| which we used in Appendix \ref{app_bandwidth} to select an appropriate bandwidth parameter, and which we described in detail in Section \ref{sim_A_estimation}. Thus, if one includes large bandwidths ($b>1$) that are essentially leading to the same estimates as a stationary model, this bandwidth selection procedure includes a model selection procedure between stationary and time-varying models. However, selecting a (roughly) stationary model with this procedure does not necessarily imply that the data generating process is stationary. The reason is that the procedure strikes a balance between stability of estimates and sensitivity to estimate time-varying parameters. If the sample size is low, the procedure will therefore select a stationary model even if the data generating process is time-varying.

Another possibility is to rely on information criteria such as the AIC \citep[see e.g.,][]{bringmann2018modeling}. Finally, one could construct a hypothesis test with the null hypothesis that the data generating process is stationary VAR model. This could be done by estimating a stationary VAR model on the data set at hand, and then generating $B$ time series of the same length as the original time series from this model. Then one fits a time-varying VAR model to each of those data sets and records a mean (over variables) prediction error. This way we obtain the sampling distribution of the prediction error under the null hypothesis, and we can perform a hypothesis test using the prediction error of the time-varying VAR model on the actual data as the test-statistic. We could for instance set $\alpha =  0.05$, which would mean that we would accept the time-varying model if its error is smaller than the 5\% quantile of the sampling distribution. For the data in this tutorial this leads to the rejection of the null-hypothesis, which means that the data generating mechanism is not a stationary VAR model and it is therefore more appropriate to fit a time-varying VAR model. We provide the code to reproduce this test on in the supplementary materials and Github \url{https://github.com/jmbh/tvvar_paper}.

\section{Discussion}\label{sec_discussion}

We compared the performance of GAM and kernel-smoothing (KS) based methods in combination with and without regularization in estimating time-varying VAR models in situations that are typical for psychological applications. Our simulation results allow researchers to select the best method amongst the ones we considered based on sample size and their assumptions about the sparsity of the true VAR model. In addition, we provided step-by-step tutorials for the KS based method using the R-package \emph{mgm} (Section \ref{sec_tutorial}) and for the GAM based method using the R-package \emph{tvvarGAM} (Appendix \ref{sec_tut_gam}).

Next to assessing the relative performance of different methods, our paper also provides the first overview of how many observations are roughly necessary to estimate time-varying VAR models. For the time-varying functions studied in our paper, already for $n>46$ the best time-varying method outperformed stationary methods, suggesting that time-varying methods can be applied to typical ESM data. However, it is important to keep in mind that if the sample size is low, the time-varying methods return very similar estimates as their stationary counterparts. Thus, if the true parameter function is heavily depending on time, and the sample size is small, time-varying methods will not be able to recover most of this dependency on time. 

There are several interesting avenues for future research on time-varying VAR models. First, in the present paper we focused on frequentist methods. However, time-varying VAR models can also be estimated in a Bayesian framework \citep{bvarsv_package}. It would be interesting to compare the performance of these methods to the methods presented in this paper. Second, the methods presented here could be extended to beyond the standard VAR models. Examples are mixed VAR models, which allow to jointly model variables defined on different domains \citep{haslbeck2015mgm}, unified Structural Equation Models (SEM) that allow an extension of SEM models to different domains \citep{kim2007unified}, or the graphical VAR model \citep{abegaz2013sparse}, which estimates both the VAR parameters and the residual structure $\Sigma$ (see Section \ref{methods_VAR}). In this model, identifying time-varying parameters is especially important, because spurious relations in the residual structure can be induced by time-varying parameters. Third, all methods discussed in this paper are based on the assumption that the true parameters are smooth functions of time. However, in some situations it might be more appropriate to assume different kinds of local stationarity, for example piece-wise constant functions \citep[e.g.,][]{gibberd2017regularized, bringmann2019tvcpvar}. It would be useful to make those alternative estimation methods available to applied researchers, and possibly combine them with the methods presented here. Fourth, the Gaussian kernel in the KS method could be replaced by kernels with finite domains such as the box car function, in order to improve the computational efficiency of the algorithm. Finally, in this paper we focused on the population performance of the two presented methods in a variety of settings. However, we did not discuss in detail how to select between models (for example stationary vs. time-varying) in a practical application. \cite{bringmann2018modeling} analyzed the performance of information criteria for selecting between stationary and time-varying VAR models with two variables. We believe that a conclusive discussion of different model selection strategies in a variety of realistic situations would be an important avenue for future work.

\section*{Acknowledgements}

We would like to thank Denny Borsboom, Fabian Dablander, Marie Deserno, Sacha Epskamp and Ois\'in Ryan and for their useful comments on earlier versions of this manuscript. We would like to thank Simon Wood for answering several inquiries about his R-package \emph{mgcv}. This research was supported by European Research Council Consolidator Grant no. 647209. 

\bibliographystyle{plain}
\bibliography{references}

\pagebreak

\appendix

\section{Sampling Variation around Aggregated Absolute Errors}\label{app_abserror_CIs}

In Figure \ref{fig_simA} we reported the mean absolute error, averaged over time points and iterations. These population level mean errors indicate which method has the lowest \emph{expected} error in a given scenario. However, it is also interesting to evaluate how large the population sampling variance is around the mean errors. We therefore display a version of Figure \ref{fig_simA} that includes the 25\% and 75\% quantiles of the population sampling distribution:

\begin{figure}[H]		
	\includegraphics[width=.95\linewidth]{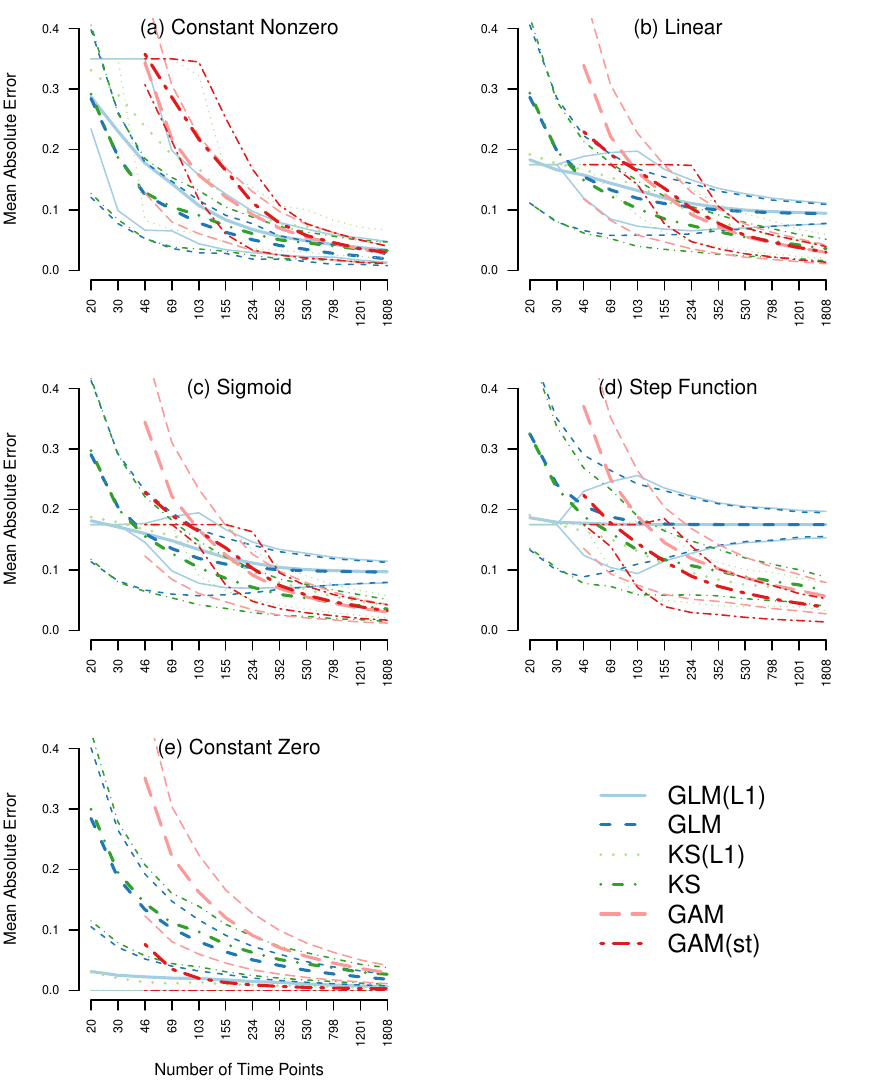}
	
	\caption{The five panels show the mean absolute estimation error (solid lines) averaged over the same type, time points, and iterations as a function of the number of observations $n$ on a log scale. We report the error of six estimation methods: stationary unregularized regression (blue), stationary $\ell_1$-regularized regression (red), time-varying  regression via kernel-smoothing (yellow), time-varying $\ell_1$-regularized regression via kernel-smoothing (green), time-varying regression via GAM (pink), and time-varying regression via GAM with thresholding at 95\% CI (orange). Some data points are missing because the respective models are not identified in that situation (see Section \ref{sim_A_estimation}). The dashed lines indicate the 25\% and 75\% quantiles, averaged over time points.}
	\label{fig_simA_app}
\end{figure}

How can we interpret these quantiles? Let's take the performance of GAM and KS for $n=103$ in panel (b) as an example. The population mean error is larger for GAM than for KS in this scenario. Note that this difference in mean errors is on the population level and therefore no test is necessary to judge its significance. However, we see that the sampling distributions of the two errors are largely overlapping. This implies that also the difference of the two errors has a large variance, which means that if $n=103$, it is difficult to predict for a specific sample whether GAM or KS has a larger error.

We see that for unregularized methods the confidence interval is large for small $n$ and becomes smaller when increasing $n$. For the $\ell_1$-regularized methods, the quantiles are first small, then increase, and then decrease again as a function of $n$. The reason is that for small $n$, these methods set all most estimates to zero, and therefore the upper and lower quantiles have the same value. An extreme case is the true zero constant function in Figure \ref{fig_simA_app} panel (e). Here both quantiles are zero for all $n$, while the mean absolute error is larger than 0 and approaches 0 with increasing $n$.

\section{Sampling Variation around Absolute Errors over Time}\label{app_abserrorOT_CIs}

\begin{figure}[H]
	\centering
	\includegraphics[width=1\linewidth]{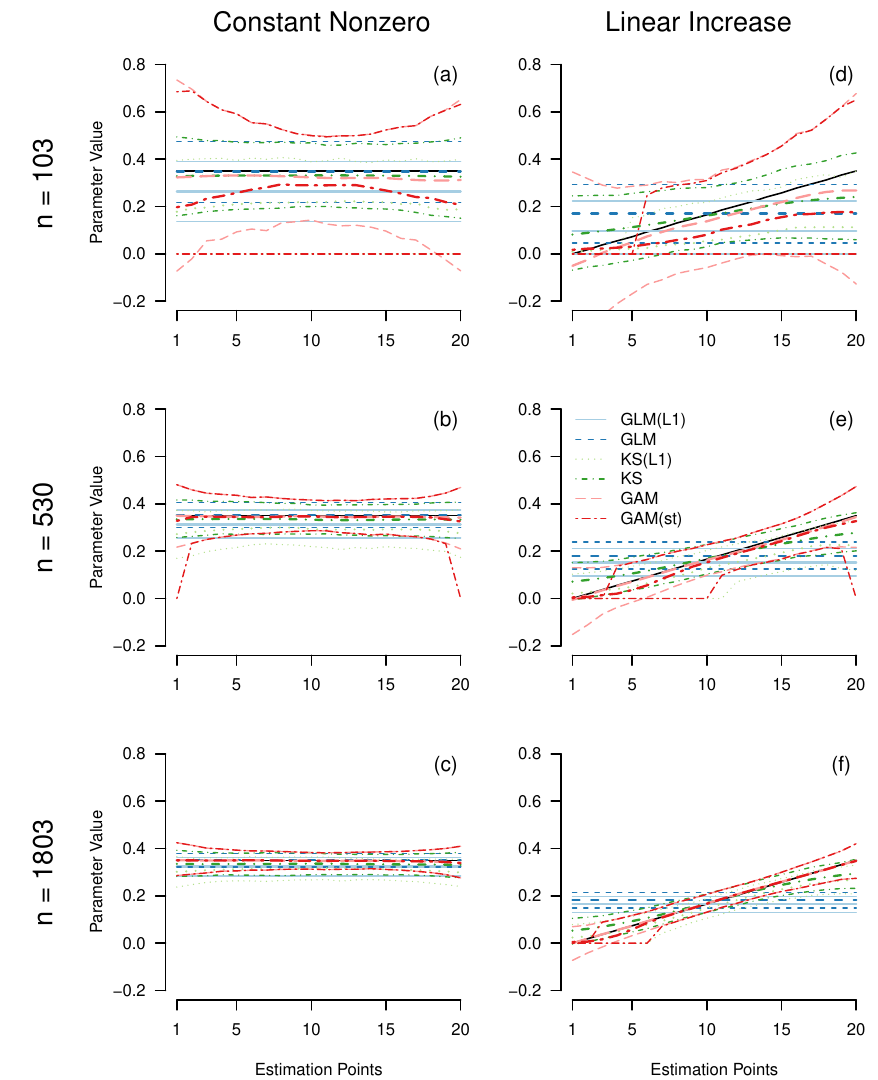}
	\caption{Mean (tick line) and standard deviations (thin line) of estimates for the constant parameter (left column), and the linear increasing parameter (right column), for $n = 103$ (top row),  $n = 530$ (second row) and $n = 1803$ (bottom row) averaged over iterations, separately for the five estimation methods: stationary $\ell_1$-regularized regression (red), unregularized regression (blue), time-varying $\ell_1$-regularized regression via kernel-smoothing (green), time-varying regression via GAM (pink), and time-varying regression via GAM with thresholding at 95\% CI (orange).}
	\label{fig_simB_wCIs}
\end{figure}

Figure \ref{fig_simB_wCIs} displays the mean estimates also shown in Figure \ref{fig_simB} in Section \ref{sim_A_results}, but in addition displays the 10\% and 90\% quantiles of the estimates. The sampling variance is small for $n=103$, but approaches zero as $n$ becomes large.

\section{Computational Cost}\label{results_comptime}

\begin{figure}[H]
	\centering
	\includegraphics[width=.70\linewidth]{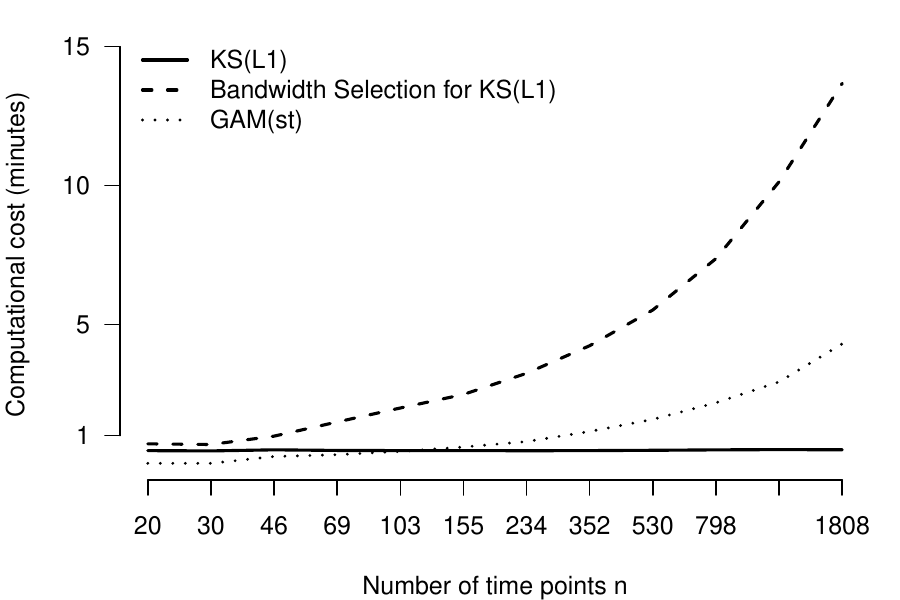}
	\caption{Computational cost in minutes to fit time-varying VAR models with the KS(L1) method (solid line) and the GAM(st) method (dotted line) as a function of observations $n$. The dashed line indicates the computational cost for selecting an appropriate bandwidth for the KS(L1) method.}
	\label{fig_simD}
\end{figure}

In Figure \ref{fig_simD} we depict the computational cost of the KS(L1) method versus the GAM(st) method. The computational complexity of the KS(L1) method is $\mathcal{O}( |E| p \log{p |L|})$, where $p$ is the number of variables, $|E|$ is the number of estimation points and $|L|$ is the number of lags included in the VAR model. The computational complexity for the bandwidth selection is $\mathcal{O}( |F| |Fs| p \log{p |L|})$, where $|F|$ is the number of folds and $|Fs|$ the number of time points in the leave-out set of each fold. For details see \cite{haslbeck2015mgm}. For the standard GAM function from the \textit{R} package \textit{mgcv} the computational complexity is $\mathcal{O}(nq^2)$, where $n$ is the number of time points modelled, and $q$ is the total number of coefficients, which increases if the number of basis functions increases \citep{wood2002gams}. Note that the credible intervals necessary for thresholding require additional computational cost. Figure \ref{fig_simD} shows the average running time (in minutes) of the two methods as a function of $n$ in the simulation reported above on a 2.60 GHz processor.

As expected, the computational cost of KS(L1) hardly increases as a function of $n$. The computational cost of GAM(st) increases roughly linear as a function of $n$. Also, the cost of the bandwidth selection scheme increases roughly linearly as a function of $n$. When considering that KS(L1) requires the data-driven selection of a bandwidth parameter, the computational cost of both method is larger for the KS(L1) method for the current setting of $p=10$ variables. However, since the computational complexity of the GAM method includes a quadratic term of the number of parameters, it is likely to perform worse when increasing the number of variables to $p > 20$. The KS(L1) method also works for huge number of variables, since its computational complexity only includes $log(p)$.

\section{Code to select Appropriate Bandwidth in KS(L1) Method}\label{app_bandwidth}

The function \verb|bwSelect()| fits time-varying VAR models with different bandwidth parameters to a set of training sets and computes the out-of-sample prediction error in the hold-out sets. We then select the bandwidth that minimizes this prediction error across variables and hold-out sets. For details about how these training/test sets are chosen exactly see \verb|?bwSelect| or \cite{haslbeck2015mgm}.

Since we fit the time-varying VAR model of our choice repeatedly, we provide all parameters we specified to the estimation function \verb|tvmvar()| as described in Section \ref{tut_KS_estimation}. In addition, we specify via \verb|bwFolds| the number of training set / test set splits, via \verb|bwFoldsize| the size of the test sets, and via \verb|bwSeq| the sequence of candidate bandwidth-values. Here, we chose ten equally spaced values in $[0.01, 1]$.

\begin{Verbatim}[commandchars=\\\{\}, fontsize=\small]
bwSeq <- seq(0.01, 1, length = 10)

set.seed(1)
bw_object <- bwSelect(data = mood_data,
                      type = rep("g", 12),
                      level = rep(1, 12),
                      bwSeq = bwSeq,
                      bwFolds = 1,
                      bwFoldsize = 20,
                      modeltype = "mvar",
                      lags = 1,
                      scale = TRUE,
                      timepoints = time_data$time_norm,
                      beepvar = time_data$beepno,
                      dayvar = time_data$dayno,
                      pbar = TRUE)

bandwidth <- bwSeq[which.min(bw_object$meanError)]

[1] 0.34
\end{Verbatim}

\begin{figure}[H]
	\centering
	\includegraphics[width=.75\linewidth]{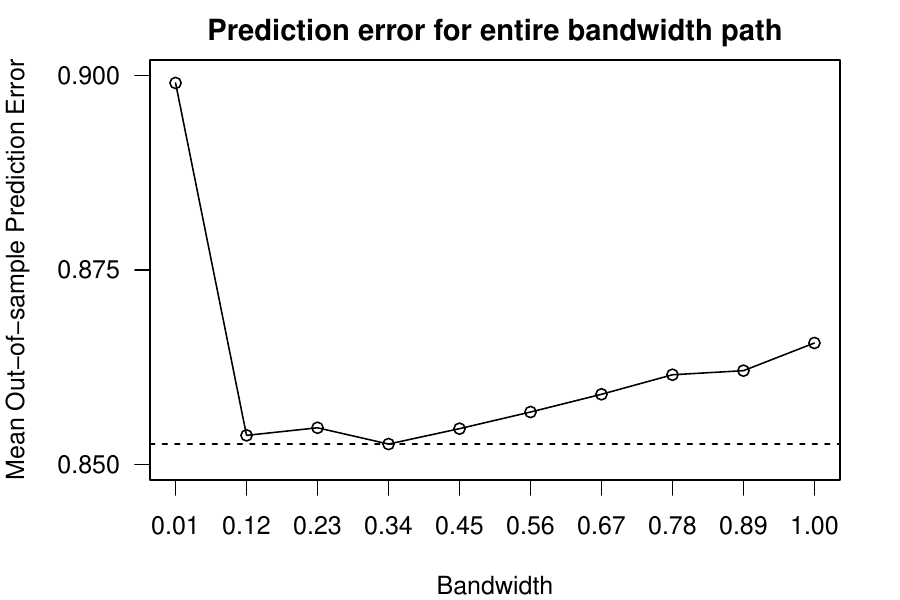}
	\caption{Average out-of-sample prediction error for different bandwidth values obtained from the function. The bandwidth value $0.34$ returns the smallest error, indicated by the dashed line.}
	\label{fig_appendix_bandwidth}
\end{figure}

The output object \verb|bw_object| contains all fitted models and unaggregated prediction errors. We see that the bandwidth $0.34$ minimized the average out-of-sample prediction error. The full bandwidth path is shown in Figure \ref{fig_appendix_bandwidth}.

The bandwidth value of $0.01$ is clearly too small, indicated by a large prediction error. The error then tends to become smaller as a function of $b$ until its minimum at $0.34$ and then increases again. Note that if the smallest/largest considered bandwidth value minimizes the error, another search should be conducted with smaller/larger bandwidth values.

\section{Estimating time-varying VAR model via GAM(st)}\label{sec_tut_gam}

Here we show how to estimate a time-varying VAR model via the GAM(st) method. All analyses are performed using the R-package \textit{tvvarGAM} \citep{tvvarGAM} and the shown code is fully reproducible, which means that the reader can execute the code while reading. The code below can also be found in an R-file on Github: \url{https://github.com/jmbh/tvvar_paper}.

\subsection{Load R-packages and dataset}

Similar to Section \ref{tutorial_mgm_data} we load the dataset from the \textit{mgm} package, and subset the 12 mood related variables. In addition, we load the \textit{tvvarGAM} package (version 0.1.1).

\begin{Verbatim}[commandchars=\\\{\}, fontsize=\small]
library(mgm) # Version 1.2-8

mood_data <- as.matrix(symptom_data$data[, 1:12]) # Subset variables
mood_labels <- symptom_data$colnames[1:12] # Subset variable labels
colnames(mood_data) <- mood_labels
time_data <- symptom_data$data_time

# Install from Github:
library(devtools)
install_github("LauraBringmann/tvvarGAM")
library(tvvarGAM)
\end{Verbatim}

\subsection{Estimating time-varying VAR model}

We use the function \verb|tvvarGAM()| to estimate the time-varying VAR model. We provide the data via the \verb|data| argument and provide an integer vector of length $n$ indicating the successiveness of measurements by specifying the number of the recorded notification and the day number via the arguments \verb|beepvar| and \verb|dayvar|. The latter is used similarly as in the \textit{mgm} package to compute the VAR design matrix. Via the argument \verb|nb| we specify the number of desired basis functions (see Section \ref{methods_GAM}). First, we estimated the model with 10 basis functions. However, because some of the edf of the smooth terms were close to 10, we doubled the number of basis functions (see discussion in Section \ref{methods_GAM}). 

\begin{Verbatim}[commandchars=\\\{\}, fontsize=\small]
tvvargam_obj <- tvvarGAM(data = mood_data, 
                         nb = 20, 
                         beepvar = time_data$beepno,
                         dayvar = time_data$dayno,
                         estimates = TRUE,
                         plot = FALSE)
\end{Verbatim}

\noindent The output object consists of a list with three entries:\\
 \verb|tvvargam_obj$Results_GAM$Estimate| is a $(p + 1) \times p \times timepoints$ array that contains the parameter estimate at each time point. The first row contains the estimated intercepts. The two other list entries have the same dimensions and contain the 5\% and 95\% confidence intervals for the estimates in \\ \verb|tvvargam_obj$Results_GAM$Estimate|. Thus, in case of the \textit{tvvarGAM} package no separate resampling scheme is necessary in order to get a measure for the reliability of parameters.

\subsection{Visualize time-varying VAR model}

\begin{figure}[H]
	\includegraphics[width=1\linewidth]{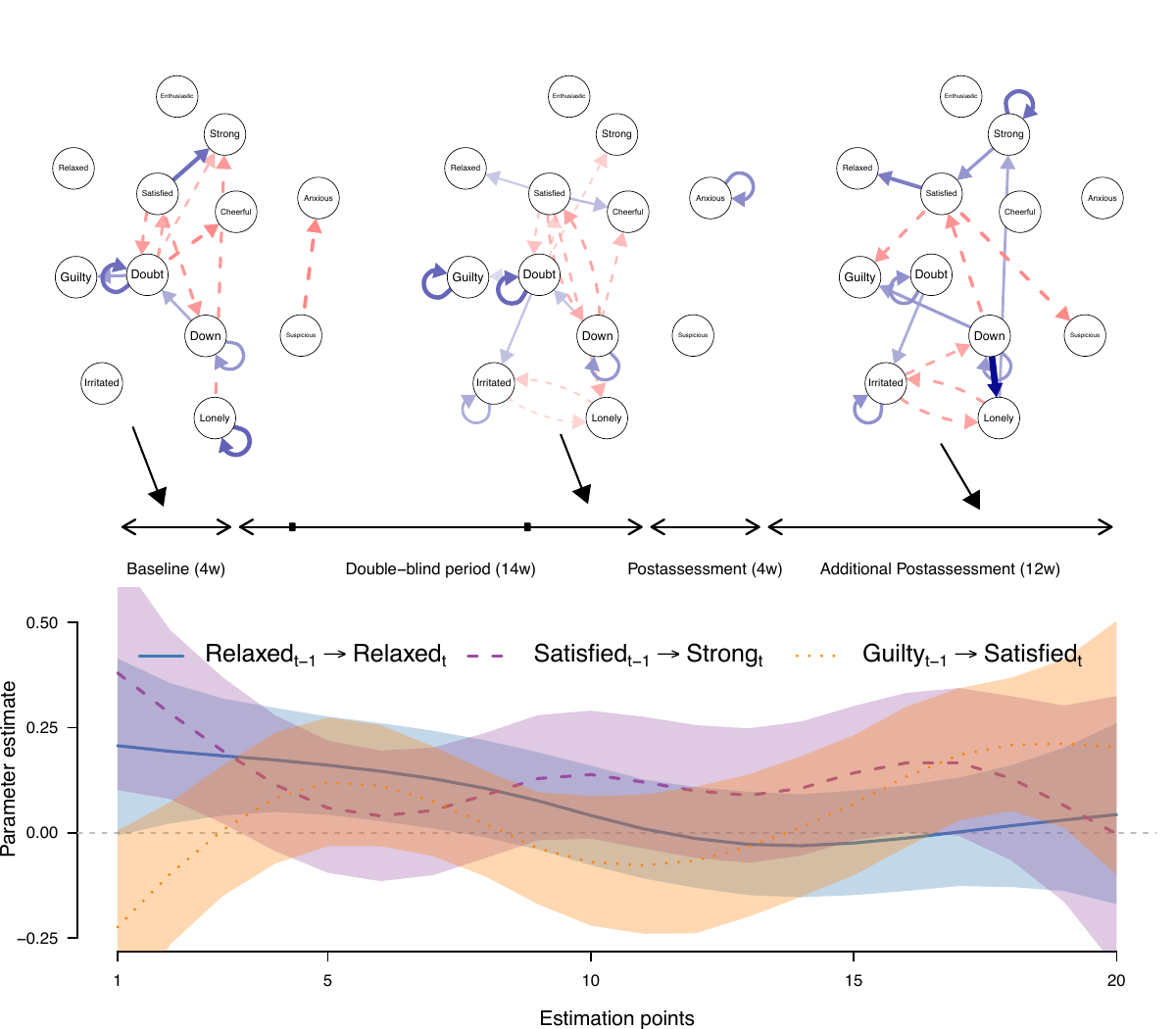}
	\caption{Top row: visualization of thresholded VAR models at estimation points 2, 10 and 18, estimated with the spline-based method. Blue arrows indicate positive relationships, red arrows indicate negative relationships, and the width of the arrows is proportional to the absolute value of the corresponding parameter. The self-loops indicate autocorrelations. Bottom row: three parameters plotted as a function of time; the points are unthresholded point estimates, the shading indicates the 5\% and 95\% credible intervals at each estimation point. }
	\label{fig_app_tvvarGAM}
\end{figure}

Figure \ref{fig_app_tvvarGAM} visualizes the part of the time-varying VAR like Figure \ref{fig_app_mgm} above, however, now with the estimates from the  \textit{tvvarGAM} package. Notice that for visualization purposes we used the tresholded version of the time-varying VAR, thus showing only the arrows that are significant (\textit{p}-value $<0.05$).

Similarly to the analysis performed with the KS(L1) method we visualize the VAR parameters at estimation points 2, 10 and 18 (top row Figure \ref{fig_app_tvvarGAM}. We see that less edges are present than in the results of the KS(L1) method, which indicates that the GAM(ks) method is more conservative. The bottom row of Figure \ref{fig_app_tvvarGAM} shows a line plot of the same three parameters as in the analysis with the KS(L1) method. We see that the effect of Relaxed on itself tends to decrease over the measured time interval, which is consistent with the results of the KS(L1) method. However, results of the cross-lagged effects of Satisfied on Strong, and of Guilty on Satisfied are only consistent with the results of the KS(1) method in the middle of the time series. The largest difference between the two methods is the increase of the effect of Guilty on Satisfied is noteworthy, while the KS(L1) method estimates a decrease. It seems that the GAM(st) estimates in the second half of the time series are incorrect, because because if one splits the time series in half and estimates two unregularized stationary VAR models, then the effect of Guilty on Satisfied is clearly negative in the second half of the time series. In general, the large changes and the much larger credible intervals at the beginning and the end of the time series indicate that the estimates are very unstable in those regions. This is consistent with the high standard deviation of estimates of the GAM and GAM(st) method shown in Figure  \ref{fig_simB_wCIs}. The code to fully reproduce Figure \ref{fig_app_tvvarGAM} is not shown here due to its length, but can be obtained from Github \url{https://github.com/jmbh/tvvar_paper}.

\end{document}